\documentclass[journal]{IEEEtran}
\ifCLASSINFOpdf
\else
   \usepackage{graphicx}
   \graphicspath{{../eps/}}
   \DeclareGraphicsExtensions{.eps}
\fi

\usepackage{amsmath}
\usepackage{amsfonts,amssymb}
\usepackage{amssymb}
\usepackage{wrapfig}
\usepackage{psfrag}
\usepackage{epstopdf}
\usepackage{cite}
\usepackage{graphicx}
\usepackage{subfigure}
\usepackage{threeparttable}
\usepackage{cases}
\usepackage{subeqnarray}
\usepackage{color}
\usepackage{underscore}
\usepackage{verbatim}
\usepackage{bm}
\usepackage{stfloats}
\usepackage{xpatch}
\newtheorem{theorem}{Theorem}

\newtheorem{algorithm}{Algorithm}
\newtheorem{proposition}{Proposition}

\usepackage{algorithm}
\usepackage{algorithmic}

\makeatletter

\begin{document}
\title{Single-Carrier Delay Alignment Modulation for Multi-IRS Aided Communication}
%
%
%
\author{
        Haiquan~Lu,
        Yong~Zeng,~\IEEEmembership{Senior Member,~IEEE,}
        Shi~Jin,~\IEEEmembership{Senior Member,~IEEE,}
        and
        Rui~Zhang,~\IEEEmembership{Fellow,~IEEE}
\thanks{This work was supported in part by the National Key R\&D Program of China with Grant number 2019YFB1803400; in part by the Natural Science Foundation of China under Grant 62071114; in part by the Fundamental Research Funds for the Central Universities 2242022k60004; in part by the Ministry of Education, Singapore under Award T2EP50120-0024; in part by the Advanced Research and Technology Innovation Centre (ARTIC) of National University of Singapore under Research Grant R-261-518-005-720, and The Guangdong Provincial Key Laboratory of Big Data Computing. (\emph{Corresponding author: Yong Zeng.}) }
\thanks{Haiquan Lu, Yong Zeng, and Shi Jin are with the National Mobile Communications Research Laboratory and Frontiers Science Center for Mobile Information Communication and Security, Southeast University, Nanjing 210096, China. Haiquan Lu and Yong Zeng are also with the Purple Mountain Laboratories, Nanjing 211111, China (e-mail: \{haiquanlu, yong_zeng, jinshi\}@seu.edu.cn). }
\thanks{Rui Zhang is with School of Science and Engineering, Shenzhen Research Institute of Big Data, The Chinese University of Hong Kong, Shenzhen, Guangdong 518172, China (e-mail: rzhang@cuhk.edu.cn). He is also with the Department of Electrical and Computer Engineering, National University of Singapore, Singapore 117583 (e-mail: elezhang@nus.edu.sg).}

}

\maketitle

\begin{abstract}
Delay alignment modulation (DAM) is a promising technology to achieve inter-symbol interference (ISI)-free single-carrier communication, by leveraging \emph{delay compensation} and \emph{path-based beamforming}, rather than the conventional channel equalization or multi-carrier transmission. In particular, when there exist a few strong time-dispersive channel paths, DAM is able to effectively align different propagation delays and achieve their constructive superposition, thus especially appealing for intelligent reflecting surfaces (IRSs)-aided communications with controllable multi-paths. In this paper, we apply single-carrier DAM to multi-IRS aided communication and study its design and achievable performance. We first provide an asymptotic analysis showing that when the number of base station (BS) antennas is much larger than the number of IRSs, an ISI-free channel can be established from the BS to the user with appropriate delay pre-compensation and the simple path-based maximal-ratio transmission (MRT) beamforming. We then consider the general system setup and study the problem of joint path-based beamforming design at the BS and phase shifts design at the IRSs for DAM transmission, by considering the three classical beamforming techniques on a per-path basis, namely the low-complexity path-based MRT beamforming to maximize the desired signal power, the path-based zero-forcing (ZF) beamforming for ISI-free DAM communication, and the optimal path-based minimum mean-square error (MMSE) beamforming to maximize the signal-to-interference-plus-noise ratio (SINR). As a comparison, orthogonal frequency-division multiplexing (OFDM)-based multi-IRS aided communication is considered for benchmarking. Simulation results are provided which demonstrate the significant performance gain of DAM over OFDM, in terms of spectral efficiency and bit error rate (BER), as well as its lower peak-to-average-power ratio (PAPR).
\end{abstract}


\IEEEpeerreviewmaketitle
\section{Introduction}
 \emph{Delay alignment modulation} (DAM) was recently proposed as a promising technique to address the inter-symbol interference (ISI) issue without relying on conventional channel equalization or multi-carrier transmission \cite{lu2022delay}. The key idea of DAM is to exploit the super spatial resolution of large antenna arrays \cite{bjornson2019massive,lu2022communicating} and multi-path channel sparsity of high-frequency channels in e.g., millimeter wave (mmWave) and TeraHertz (THz) communications \cite{akdeniz2014millimeter,zeng2016millimeter,rappaport2019wireless,chen2021hybrid}, for performing \emph{delay pre/post-compensation} \cite{lu2022delay,zeng2018multi} and \emph{path-based beamforming}. Specifically, by deliberately introducing symbol delays at the transmitter to match the channel's multi-path delays, together with appropriate path-based beamforming, all multi-path signal components may arrive simultaneously and constructively at the receiver, without the detrimental ISI. In general, DAM is able to manipulate the channel delay spread with simple spatial-delay processing, which provides an effective approach to combat the time-dispersive channel for more efficient single- or multi-carrier transmissions \cite{lu2022Manipulating}. For example, by integrating DAM with orthogonal frequency-division multiplexing (OFDM), a novel DAM-OFDM technique was proposed in \cite{lu2022Manipulating}, which provides a framework that unifies single-carrier and multi-carrier transmissions by flexibly adjusting the number of sub-carriers and cyclic prefix (CP) overhead, while guaranteeing ISI-free communications without the need of sophisticated channel equalizations. Efficient channel estimation method for DAM was studied in \cite{ding2022channel}, and DAM-based integrated sensing and communication (ISAC) was investigated in \cite{xiao2022integrated}.

 Compared to the conventional ISI mitigation techniques, such as time-domain equalization (TEQ) and OFDM communication, DAM has the following significant differences and advantages. In contrast to TEQ techniques such as channel shortening \cite{melsa1996impulse,martin2005unification} and time-reversal (TR) \cite{emami2004matched,han2012time,Why2016Chen} that mainly rely on the time-domain processing, DAM addresses the ISI issue via joint spatial-delay processing by fully leveraging the super spatial resolution of large antenna arrays and multi-path sparsity of high-frequency channels. Note that TR essentially treats the multi-path channel as a matched filter \cite{emami2004matched,han2012time,Why2016Chen,pitarokoilis2012optimality,pitaval2021channel}, and the ISI is alleviated by relying on the rate back-off technique  \cite{emami2004matched,han2012time,Why2016Chen}, or eliminated only when the number of base station (BS) antennas goes to infinity \cite{pitarokoilis2012optimality,pitaval2021channel}. By contrast, DAM is able to completely eliminate the ISI as long as the number of antennas is no smaller than that of multi-paths, via the flexible delay compensation and path-based beamforming \cite{lu2022delay}. On the other hand, unlike multi-carrier transmission such as OFDM, the single-carrier-based DAM inherently circumvents the practical drawbacks of OFDM, such as high peak-to-average-power ratio (PAPR), sensitivity to carrier frequency offset (CFO), and the severe out-of-band (OOB) emission \cite{goldsmith2005wireless,heath2018foundations}. Although various techniques have been studied to resolve those issues in OFDM, such as amplitude clipping and discrete Fourier transform-spread OFDM (DFT-s-OFDM) for PAPR reduction \cite{han2005overview,tong20216g}, windowing and filtering for OOB emission suppression \cite{tong20216g,nissel2017filter}, as well as CFO estimation and compensation \cite{yao2005blind}, they either incur performance degradation or aggravate the signal processing complexity. Besides, DAM only requires one guard interval for each channel coherence block, rather than one CP for each OFDM symbol, which renders DAM more spectrally efficient than OFDM \cite{lu2022delay,lu2022Manipulating}. Lastly, it is worth remarking that DAM can be naturally combined with existing single- or multi-carrier ISI techniques by flexibly manipulating the channel delay spread with spatial-delay processing, as revealed in \cite{lu2022Manipulating}.

 More recently, a novel multi-carrier modulation technique termed orthogonal time frequency space (OTFS) was proposed \cite{hadani2017orthogonal}. While OTFS is quite promising in high-mobility scenarios, it requires high signal processing complexity due to the large-dimension equivalent channel \cite{xiao2021overview}. Furthermore, the signal detection of OTFS has to be performed across an entire OTFS frame, which leads to high signal processing latency for a large OTFS frame. By contrast, the equalization-free DAM transmission enables instantaneous signal detection, making it quite appealing for ultra-reliable low-latency communication (URLLC) in the fifth-generation (5G)/sixth-generation (6G) wireless networks. It is worth mentioning that a technique similar to DAM was studied in
 \cite{taniguchi2004maximum}, where tapped delay line structure was used to alleviate the channel frequency selectivity. However, the work \cite{taniguchi2004maximum} did not exploit the high spatial resolution nor multi-path sparsity, and the ISI cannot be completely eliminated in general. Furthermore, no rigorous performance analysis and signal design were provided in \cite{taniguchi2004maximum}.

 Note that DAM is most effective when there exist a few strong channel multi-paths. Besides high-frequency systems such as mmWave or THz communications that have the intrinsic multi-path sparsity, another scenario with limited significant multi-paths is intelligent reflecting surface (IRS)-aided communication, which has received dramatic attention recently \cite{wu2019towards,wu2021intelligentTutorial,basar2019wireless,han2019large,he2020cascaded,
 lu2021aerial,feng2021wireless,di2020smart,Tang2021WirelessCM,
 huang2019reconfigurable,wu2019intelligent,pan2022overview,
 shao2022target,han2020cooperative,you2021wireless,gao2021distributed,
 ding2023analysis,jia2022robust}. However, most existing works on IRS-aided communication focus on narrowband frequency-flat communications without considering the ISI issue. For example, by considering the double-reflection cascaded channel, the theoretical power scaling and channel estimation scheme for cooperative double-IRS aided communications were studied in \cite{han2020cooperative} and \cite{you2021wireless}, respectively. Moreover, to reduce the complexity of instantaneous channel state information (CSI)-adaptive phase shift design, the quasi-static phase shift designs based on statistical CSI were considered for multi-IRS aided narrowband communications in \cite{gao2021distributed,ding2023analysis}. While for wideband communications, the time-dispersive channel caused by multi-path propagation becomes a major impairment, for which the detrimental ISI needs to be considered. Some efforts have been devoted to IRS-aided wideband communications using OFDM. For example, in \cite{Yang2020IntelligentRS,zheng2020intelligent}, the practical transmission protocols were proposed for IRS-aided OFDM systems. The extensions to the cooperative multi-IRS aided multiple-input single-output (MISO)-OFDM and the IRS-aided multiple-input multiple-output (MIMO)-OFDM wideband communications were studied in \cite{he2021cooperative} and \cite{zhang2020capacity,li2021intelligent}, respectively.

 Intuitively, the strong virtual line-of-sight (LoS) paths created by the smart reflection of IRSs render DAM especially appealing for IRS-aided communications. Therefore, in this paper, we study the multi-IRS aided communication where different IRSs create multi-paths with different delays, and single-carrier DAM is applied for mitigating the ISI. We first provide performance analysis for DAM by assuming that the number of BS antennas, $N_t$, is much larger than the number of IRSs/multi-paths, $L$. Then, for the general system setup with arbitrary values of $N_t$ and $L$, we study the problem of joint path-based beamforming design at the BS and phase shifts design at the IRSs for DAM transmission. In particular, three classical beamforming techniques on the per-path basis are investigated, i.e., path-based zero-forcing (ZF), minimum mean-square error (MMSE), and maximal-ratio transmission (MRT). The main contributions of this paper are summarized as follows.
 \begin{itemize}[\IEEEsetlabelwidth{12)}]
 \item Firstly, we present the input-output relationship of single-carrier DAM for multi-IRS aided communication systems. For the ideal case with ${N_t} \gg L$, by exploiting the asymptotically orthogonal property among channels associated with distinct multi-path delays, we show that an ISI-free additive white Gaussian noise (AWGN) channel can be attained between the BS and the user with appropriate delay pre-compensation and the simple path-based MRT beamforming. Moreover, the optimal beamforming vectors and IRS phase shifts to maximize the signal-to-noise ratio (SNR) are derived. In addition, the centralized versus distributed multi-IRS deployment strategies are compared based on the closed-from SNR expressions.
 \item Secondly, for any finite BS antenna number $N_t$ and IRS number $L$, the three classical beamforming design approaches, namely ZF, MMSE, and MRT, are respectively studied, on the per-path basis to cater to the DAM transmission. Specifically, provided that ${N_t} \ge L+1$, the time-dispersive channel created by multiple IRSs can be transformed into an ISI-free AWGN channel with path-based ZF beamforming, without relying on the conventional channel equalization or multi-carrier transmission, thanks to the perfect delay alignment. On the other hand, when perfect delay alignment is infeasible or undesirable, the optimal path-based MMSE beamforming and the low-complexity path-based MRT beamforming are designed to maximize the signal-to-interference-plus-noise ratio (SINR) and the desired signal power, respectively, by jointly optimizing the path-based beamforming vectors and IRS phase shifts.
 \item Lastly, as a comparison, we consider the benchmarking scheme of OFDM-based multi-IRS aided communication, and the impact of guard interval overhead for single-carrier DAM and OFDM is analyzed. Simulation results demonstrate that, thanks to the saving of guard interval overhead, DAM outperforms OFDM in terms of spectral efficiency and bit error rate (BER), and also achieves lower PAPR.
\end{itemize}

 The rest of this paper is organized as follows. Section \ref{sectionSystemModel} introduces the system model and provides the asymptotic analysis for DAM communication. In Sections \ref{sectionPathBasedZFBeamforming}-\ref{sectionPathBasedMRTBeamforming}, the joint design of path-based beamforming at the BS and phase shifts at the IRSs are studied by considering the path-based ZF, MMSE, and MRT beamforming, respectively. Section \ref{sectionBenchmarkOFDMScheme} presents the benchmarking scheme of OFDM-based multi-IRS aided communication. Simulation results are presented in Section \ref{sectionNumericalResults}, and this paper is concluded in Section \ref{sectionConclusion}.

 \emph{Notations:} Scalars are denoted by italic letters. Vectors and matrices are denoted by bold-face lower- and upper-case letters, respectively. ${{\mathbb{C}}^{M \times N}}$ denotes the space of $M \times N$ complex-valued matrices. For an arbitrary-size matrix ${\bf A}$, its complex conjugate, transpose, and Hermitian transpose are denoted by ${\bf A}^*$, ${\bf A}^T$, and ${\bf A}^H$, respectively. For a vector $\bf{x}$, ${\left\| {\bf{x}} \right\|_1}$ and $\left\| {\bf{x}} \right\|$ denote its ${\ell}_1$-norm and ${\ell}_2$-norm, respectively, and ${\rm{arg}}\left( {\bf{x}} \right)$ denotes a vector consisting of the phases of the corresponding elements in $\bf x$. The notations $\circledast$ and $\otimes$ denote the linear convolution and the Kronecker product operations, respectively. The distribution of a circularly symmetric complex Gaussian (CSCG) random vector with mean $\bf{x}$ and covariance matrix $\bf{\Sigma}$ is denoted by ${\cal CN}\left( {\bf{x},\bf{{\Sigma}}} \right)$; and $\sim$ stands for ``distributed as". The symbol $j$ denotes the imaginary unit of complex numbers, with ${j^2} =  - 1$. ${\mathbb E}\left({\cdot}\right)$ denotes the statistical expectation. ${\mathop{\rm Re}\nolimits} \left\{ {\cdot} \right\}$ denotes the real part of a complex number. ${\cal O}\left({\cdot}\right)$ denotes the standard big-O notation.

\section{System Model And Asymptotic Analysis}\label{sectionSystemModel}
\subsection{System Model}\label{subsectionSystemModel}
 \begin{figure}[!t]
 \centering
 \centerline{\includegraphics[width=3.25in,height=2.4in]{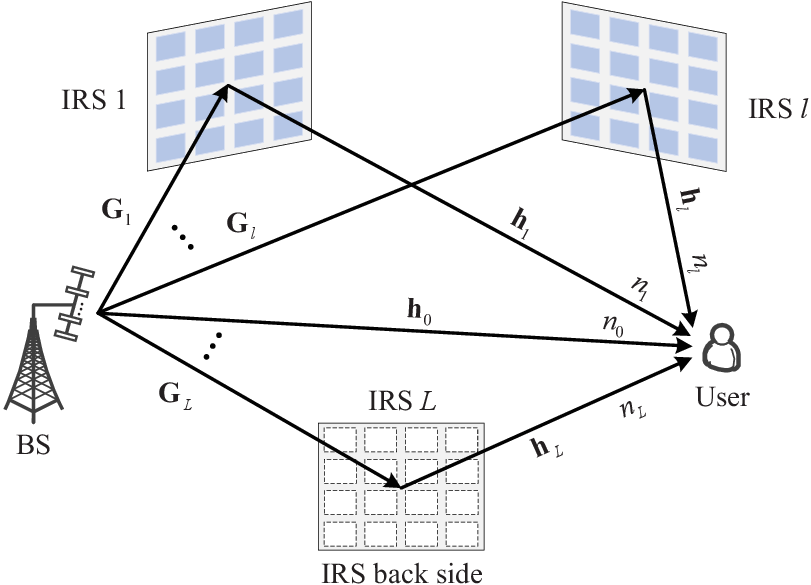}}
 \caption{A multi-IRS aided communication system.}
 \label{systemModel}
 \end{figure}
 As shown in Fig.~\ref{systemModel}, we consider a multi-IRS aided wideband communication system, where $L$ IRSs are deployed to assist in the communication from a BS to a user. The BS is equipped with $N_t$ antennas, and the user has one antenna. Each IRS consists of $M = {M_h}{M_v}$ passive reflecting elements, where ${M_h}$ and ${M_v}$ denote the number of elements in the horizontal and vertical dimensions, respectively. Denote by ${{\bf{\Theta }}_l} = {\rm{diag}}\left( {{e^{j{\theta _{l,1}}}}, \cdots ,{e^{j{\theta _{l,m}}}}, \cdots ,{e^{j{\theta _{l,M}}}}} \right)$ the diagonal phase-shift matrix of IRS $l$, where ${\theta _{l,m}} \in \left[ {0,2\pi } \right)$ is the phase shift of the $m$th reflecting element of IRS $l$, with $m =  {1, \cdots, M} $ and $l = {1, \cdots, L}$. The direct channel from the BS to the user without the reflection by any IRS is denoted by ${\bf{h}}_0^H \in {{\mathbb C}^{1 \times {N_t}}}$. Let ${{\bf{G}}_l} \in {{\mathbb C}^{M \times {N_t}}}$ denote the channel matrix from the BS to IRS $l$, and ${\bf{h}}_l^H \in {{\mathbb C}^{1 \times M}}$ denote the channel vector from IRS $l$ to the user. We focus on the time-dispersive multi-path fading channel between the BS and the user induced by multiple IRSs, where the strongest channel tap is considered for each pair of nodes. Such a time-dispersive channel model is practically valid due to the following reasons. On one hand, for high-frequency systems such as mmWave communications, the effect of LoS component typically dominates since the high frequency signals usually suffer from severe scattering and diffraction losses \cite{akdeniz2014millimeter}. As such, the strongest direct channel tap between the BS and the user is considered. On the other hand, to fully exploit the benefits of IRSs, it is desirable to deploy them in LoS conditions with the BS/user \cite{wu2021intelligentTutorial}. Moreover, the signals experiencing more than one reflection are ignored since they are typically much weaker than those with one reflection only \cite{wu2019intelligent,zhang2020capacity}. We assume quasi-static block fading model, where the channel remains unchanged within each channel coherence interval. The discrete-time equivalent of the channel impulse response from the BS to the user can be expressed as
 \begin{equation}\label{discreteTimeChannelImpulseResponse}
 {\bf{h}}_{{\rm{eq}}}^H\left[ n \right] = {\bf{h}}_0^H\delta \left[ {n - {n_0}} \right] + \sum\limits_{l = 1}^L {{\bf{h}}_l^H{{\bf{\Theta }}_l}{{\bf{G}}_l}\delta \left[ {n - {n_l}} \right]},
 \end{equation}
 where $n_0$ denotes the discrete-time delay of the direct channel, and $n_l$ denotes the delay of the reflected channel via IRS $l$. For ease of exposition, we assume that different IRSs create multi-paths with different resolvable delays, i.e., ${n_l} \ne  {n_{l'}}$, $\forall l\ne l'$. Let ${n_{\min }} \triangleq \mathop {\min }\limits_{0 \le l \le L} {n_l}$ and ${n_{\max }} \triangleq \mathop {\max }\limits_{0 \le l \le L} {n_l}$ be the minimum and maximum delay over all the $L+1$ multi-paths, respectively. Then the channel delay spread is defined as ${n_{\rm{span}}} \triangleq {n_{\max }} - {n_{\min }}$. To reveal the fundamental performance limit achieved by DAM, we assume that perfect CSI is available at the BS. In particular, the CSI acquisition of the individual path information, e.g., angle of arrival/departure (AoA/AoD) and propagation delay will become more feasible with the use of large antenna arrays at the BS \cite{bjornson2019massive,lu2022communicating} and the trend of integrated sensing and communication (ISAC) \cite{tong20216g,zhang2021enabling}. A preliminary study on CSI acquisition for DAM is being pursued in another parallel work \cite{ding2022channel}, and the channel estimation for the considered multi-IRS system requires more in-depth studies in the future.

 Denote by ${\bf{x}}\left[ n \right] \in {{\mathbb{C}}^{{N_t} \times 1}}$ the discrete-time equivalent of the transmitted signal. Then the received signal is
 \begin{equation}\label{receiveSignal}
 \begin{aligned}
 y\left[ n \right] &= {\bf{h}}_{{\rm{eq}}}^H\left[ n \right] \circledast {\bf{x}}\left[ n \right] + z\left[ n \right]\\
 &= {\bf{h}}_0^H{\bf{x}}\left[ {n - {n_0}} \right] + \sum\limits_{l = 1}^L {{\bf{h}}_l^H{{\bf{\Theta }}_l}{{\bf{G}}_l}{\bf{x}}\left[ {n - {n_l}} \right]}  + z\left[ n \right],
 \end{aligned}
 \end{equation}
 where $z\left[ n \right] \sim {\cal CN}\left( {0,{\sigma ^2}} \right)$ is the AWGN.

 Before introducing the DAM transmission and providing its asymptotic performance analysis for multi-IRS aided wideband communication, we first investigate the correlation of the channels associated with the $L+1$ multi-paths. For ease of illustration, we assume here the LoS-dominating channel between the BS and each IRS $l$, and denote by ${\varphi _l}$ and ${\phi _l}$ the AoD and AoA from the BS to IRS $l$, respectively. The channel matrix ${\bf{G}}_l$ is then expressed as ${{\bf{G}}_l} = {\alpha _l}{{\bf{a}}_R}\left( {{\phi _l}} \right){\bf{a}}_T^H\left( {{\varphi _l}} \right)$, where ${\alpha _l}$ is the complex-valued path gain, and ${{\bf{a}}_R}\left( {{\phi _l}} \right) \in {{\mathbb C}^{M \times 1}}$ and ${{\bf{a}}_T}\left( {{\varphi _l}} \right) \in {{\mathbb C}^{{N_t} \times 1}}$ denote the receive and transmit array response vectors, respectively. If the BS is equipped with the basic uniform linear array (ULA) with adjacent elements separated by half wavelength, the transmit array response vector can be expressed as \cite{tse2005fundamentals}
 \begin{equation}\label{transmitArrayResponse}
 {{\bf{a}}_T}\left( {{\varphi _l}} \right) = {\left[ {1,{e^{ - j\pi \cos {\varphi _l}}}, \cdots ,{e^{ - j\pi \left( {{N_t} - 1} \right)\cos {\varphi _l}}}} \right]^T}.
 \end{equation}
 Similarly, the direct channel ${\bf h}_0$ can be expressed as ${{\bf{h}}_0} = {\alpha _0}{{\bf{a}}_T}\left( {{\varphi _0}} \right)$, where ${\alpha _0}$ is the complex coefficient of the direct channel, and ${{\varphi _0}}$ is its AoD. It then follows from \cite{lu2022Manipulating} that as long as the $L$ IRSs and the direct link are associated with distinct AoDs, the corresponding transmit array response vectors tend to be asymptotically orthogonal when ${N_t} \gg L$, i.e.,
 \begin{equation}\label{asymptoticalProperty}
 \frac{{\bf{a}}_T^H\left( {{\varphi _{l}}} \right){{\bf{a}}_T}\left( {{\varphi _{l'}}} \right)}{N_t} \to 0, \ \forall l \ne l'.
 \end{equation}
 Furthermore, since ${{\bf{G}}_l} = {\alpha _l}{{\bf{a}}_R}\left( {{\phi _l}} \right){\bf{a}}_T^H\left( {{\varphi _l}} \right)$, based on \eqref{asymptoticalProperty}, we have
 \begin{equation}\label{asymptoticalProperty2}
 \frac{{{{\bf{G}}_l}{\bf{G}}_{l'}^H}}{{{N_t}}} \to {\bf{0}},\ \forall l \ne l',\ l,l' \in \left[1,L \right],
 \end{equation}
  \begin{equation}\label{asymptoticalProperty3}
 \frac{{{{\bf{G}}_l}{{\bf{h}}_0}}}{{{N_t}}} \to {\bf{0}},\ \forall l \in \left[ {1,L} \right].
 \end{equation}
 As discussed in \cite{lu2022Manipulating} and further elaborated in Section \ref{subsectionKeyIdeaOfDAM}, such a property motivates our proposed DAM technique as a new solution for ISI-free communication, without relying on sophisticated channel equalization or multi-carrier transmission.

 \subsection{DAM and Asymptotic Analysis}\label{subsectionKeyIdeaOfDAM}
 With single-carrier DAM, the signal transmitted by the BS is \cite{lu2022delay,lu2022Manipulating}
 \begin{equation}\label{transmitSignalDAM}
 {\bf{x}}\left[ n \right] = \sum\limits_{l' = 0}^L {{{\bf{f}}_{l'}}s\left[ {n - {\kappa _{l'}}} \right]},
 \end{equation}
 where $s\left[ n \right]$ is the independent and identically distributed (i.i.d.) information-bearing symbol with zero mean and unit variance, ${\bf{f}}_{l'} \in {{\mathbb C}^{{N_t} \times 1}}$ denotes the path-based transmit beamforming vector associated with path $l'$, and $\kappa _{l'} \ge 0$ is the deliberately introduced delay pre-compensation for the symbol sequence $s\left[ n \right]$, with ${\kappa _{l}} \ne {\kappa _{l'}}$, $\forall l \ne l'$. The transmit power of the BS is
 \begin{equation}\label{transmitPowerConstraint}
 \begin{aligned}
 {\mathbb{E}}\left[ {{{\left\| {{\bf{x}}\left[ n \right]} \right\|}^2}}\right] = \sum\limits_{l' = 0}^L {{\mathbb{E}}\left[ {{{\left\| {{{\bf{f}}_{l'}}s\left[ {n - {\kappa_{l'}}} \right]} \right\|}^2}} \right]} = \sum\limits_{l' = 0}^L {{{\left\| {{{\bf{f}}_{l'}}} \right\|}^2}}  \le P,
 \end{aligned}
 \end{equation}
 where $P$ denotes the available transmit power, and the first equality holds since $s\left[ n \right]$ is independent across different $n$ and ${\kappa _{l}} \ne {\kappa _{l'}}$, $\forall l \neq l'$. By substituting \eqref{transmitSignalDAM} into \eqref{receiveSignal}, the signal received by the user can be expressed as
 \begin{equation}\label{receiveSignalDAM}
 \hspace{-1ex}
 \begin{aligned}
 y\left[ n \right] &{\rm =} \sum\limits_{l' = 0}^L {{\bf{h}}_0^H{{\bf{f}}_{l'}}s\left[ {n - {\kappa _{l'}} - {n_0}} \right]}  {\rm +} \sum\limits_{l = 1}^L {{\bf{h}}_l^H{{\bf{\Theta }}_l}{{\bf{G}}_l}{{\bf{f}}_l}s\left[ {n - {\kappa _l} - {n_l}} \right]} \\
 &+ \sum\limits_{l = 1}^L {\sum\limits_{l'=0,l' \ne l}^L {{\bf{h}}_l^H{{\bf{\Theta }}_l}{{\bf{G}}_l}{{\bf{f}}_{l'}}s\left[ {n - {\kappa _{l'}} - {n_l}} \right]} }  + z\left[ n \right].
 \end{aligned}
 \end{equation}
 By letting ${\kappa _{l'}} = {n_{\max }} - {n_{l'}} \ge 0$, $\forall l' \in \left[0,L\right]$, we have
 \begin{equation}\label{IRSDelayCompensationReceivedSignal}
 \hspace{-1ex}
 \begin{aligned}
 &y\left[ n \right] = \underbrace {\left( {{\bf{h}}_0^H{{\bf{f}}_0} + \sum\limits_{l = 1}^L {{\bf{v}}_l^H{\rm{diag}}\left( {{\bf{h}}_l^H} \right){{\bf{G}}_l}{{\bf{f}}_l}} } \right)s\left[ {n - {n_{\max }}} \right]}_{{\rm{desired\ signal}}} +  \\
 &\underbrace {\sum\limits_{l' = 1}^L {{\bf{h}}_0^H{{\bf{f}}_{l'}}s\left[ {n - {n_{\max }} + {n_{l'}} - {n_0}} \right]} }_{{\rm{ISI}}} +  \\
 &\underbrace {\sum\limits_{l = 1}^L {\sum\limits_{l' = 0,l' \ne l}^L {{\bf{v}}_l^H{\rm{diag}}\left( {{\bf{h}}_l^H} \right){{\bf{G}}_l}{{\bf{f}}_{l'}}s\left[ {n - {n_{\max }} + {n_{l'}} - {n_l}} \right]} } }_{{\rm{ISI}}} + z\left[ n \right],
 \end{aligned}
 \end{equation}
 where we have used the equation ${\bf{h}}_l^H{{\bf{\Theta }}_l}{{\bf{G}}_l} = {\bf{v}}_l^H{\rm{diag}}\left( {{\bf{h}}_l^H} \right){{\bf{G}}_l}$, with ${{\bf{v}}_l} = {\left[ {{e^{j{\theta _{l,1}}}}, \cdots ,{e^{j{\theta _{l,M}}}}} \right]^H}$. It is observed from \eqref{IRSDelayCompensationReceivedSignal} that if the user is locked to the delay $n_{\max}$, then the first term contributes to the desired signal, while the second and the third terms are the detrimental ISI.

 Fortunately, for massive MIMO system with ${N_t} \gg L$, we will show that the ISI term in \eqref{IRSDelayCompensationReceivedSignal} vanishes with the simple path-based MRT beamforming, thanks to the asymptotically orthogonal properties \eqref{asymptoticalProperty}-\eqref{asymptoticalProperty3}. With path-based MRT beamforming, each beamforming vector ${{\bf{f}}_{l'}}$ in \eqref{transmitSignalDAM} matches the transmit array response vector of the corresponding multi-path, which is given by
 \begin{equation}\label{pathBasedMRTBeamformingAsymptotical}
 {{\bf{f}}_{l'}} = \sqrt {{p_{l'}}} \frac{{{{\bf{a}}_T}\left( {{\varphi _{l'}}} \right)}}{{\left\| {{{\bf{a}}_T}\left( {{\varphi _{l'}}} \right)} \right\|}} = \sqrt {{p_{l'}}} \frac{{{{\bf{a}}_T}\left( {{\varphi _{l'}}} \right)}}{{\sqrt {{N_t}} }},
 \end{equation}
 where $p_{l'}$ denotes the power allocation for the $l'$th multi-path. By substituting \eqref{pathBasedMRTBeamformingAsymptotical} into \eqref{IRSDelayCompensationReceivedSignal}, and after scaling the received signal $y\left[ n \right]$ by
 $1/\sqrt {{N_t}} $, we have \eqref{IRSDelayCompensationReceivedSignalMRT}, shown at the top of the next page, where we have used the asymptotical property \eqref{asymptoticalProperty}.
 \newcounter{mytempeqncnt1}
 \begin{figure*}
 \normalsize
 \setcounter{mytempeqncnt1}{\value{equation}}
 \begin{equation}\label{IRSDelayCompensationReceivedSignalMRT}
 \hspace{-1ex}
 \begin{aligned}
 \frac{1}{{\sqrt {{N_t}} }}y\left[ n \right] &= \left( {\sqrt {{p_0}} \alpha _0^* + \sum\limits_{l = 1}^L {\sqrt {{p_l}} {\alpha _l}{\bf{v}}_l^H{\rm{diag}}\left( {{\bf{h}}_l^H} \right){{\bf{a}}_R}\left( {{\phi _l}} \right)} } \right)s\left[ {n - {n_{\max }}} \right] + \sum\limits_{l' = 1}^L {\sqrt {{p_{l'}}} \alpha _0^*\frac{{{\bf{a}}_T^H\left( {{\varphi _0}} \right){{\bf{a}}_T}\left( {{\varphi _{l'}}} \right)}}{{{N_t}}}s\left[ {n - {n_{\max }} + {n_{l'}} - {n_0}} \right]}  +  \\
 &\ \ \ \ \ \sum\limits_{l = 1}^L {\sum\limits_{l' = 0,l' \ne l}^L {\sqrt {{p_{l'}}} {\alpha _l}{\bf{v}}_l^H{\rm{diag}}\left( {{\bf{h}}_l^H} \right){{\bf{a}}_R}\left( {{\phi _l}} \right)\frac{{{\bf{a}}_T^H\left( {{\varphi _l}} \right){{\bf{a}}_T}\left( {{\varphi _{l'}}} \right)}}{{{N_t}}}s\left[ {n - {n_{\max }} + {n_{l'}} - {n_l}} \right]} }  + \frac{1}{{\sqrt {{N_t}} }}z\left[ n \right]\\
 &\to \left( {\sqrt {{p_0}} \alpha _0^* + \sum\limits_{l = 1}^L {\sqrt {{p_l}} {\alpha _l}{\bf{v}}_l^H{\rm{diag}}\left( {{\bf{h}}_l^H} \right){{\bf{a}}_R}\left( {{\phi _l}} \right)} } \right)s\left[ {n - {n_{\max }}} \right] + \frac{1}{{\sqrt {{N_t}} }}z\left[ n \right].
 \end{aligned}
 \end{equation}
 \hrulefill
 \end{figure*}
 By multiplying \eqref{IRSDelayCompensationReceivedSignalMRT} with $\sqrt {{N_t}} $, the signal in \eqref{IRSDelayCompensationReceivedSignal} reduces to
 \begin{equation}\label{IRSDelayCompensationReceivedSignalMRT2}
 \begin{aligned}
 &y\left[ n \right] \to \sqrt {{N_t}} \Big( {\sqrt {{p_0}} \alpha _0^* + \sum\limits_{l = 1}^L {\sqrt {{p_l}} {\alpha _l}{\bf{v}}_l^H{\rm{diag}}\left( {{\bf{h}}_l^H} \right){{\bf{a}}_R}\left( {{\phi _l}} \right)} } \Big) \times \\
 &\ \ \ \ \ \ \ \ \ \ \ \ \ \ \ \ \ \ \ \ \ s\left[ {n - {n_{\max }}} \right] + z\left[ n \right].
 \end{aligned}
 \end{equation}
 It is observed from \eqref{IRSDelayCompensationReceivedSignal} and \eqref{IRSDelayCompensationReceivedSignalMRT2} that thanks to the asymptotical orthogonal properties \eqref{asymptoticalProperty}-\eqref{asymptoticalProperty3}, the ISI terms in \eqref{IRSDelayCompensationReceivedSignal} vanish almost surely with the proposed DAM transmission using delay pre-compensation and the simple path-based MRT beamforming. Therefore, the original time-dispersive channel has been transformed to an ISI-free AWGN channel, while still benefiting from the multiplicative gain contributed by all the $L+1$ multi-path signal components. Also, this is free from the sophisticated channel equalization or multi-carrier transmission. Note that similar derivation and conclusion have been made in \cite{lu2022Manipulating} for DAM communication without IRS. The SNR of \eqref{IRSDelayCompensationReceivedSignalMRT2} is then given by
 \begin{equation}\label{IRSMRTSNR}
 \gamma = \frac{{{N_t}{{\left| {\sqrt {{p_0}} \alpha _0^* + \sum\limits_{l = 1}^L {\sqrt {{p_l}} {\alpha _l}{\bf{v}}_l^H{\rm{diag}}\left( {{\bf{h}}_l^H} \right){{\bf{a}}_R}\left( {{\phi _l}} \right)} } \right|}^2}}}{{{\sigma ^2}}}.
 \end{equation}
 It is not difficult to see that to maximize the SNR in \eqref{IRSMRTSNR}, the phase shift vectors ${\bf{v}}_l$ should be designed so that $\arg \left( {\alpha _0^*} \right) = \arg \left( {{\alpha _l}{\bf{v}}_l^H{\rm{diag}}\left( {{\bf{h}}_l^H} \right){{\bf{a}}_R}\left( {{\phi _l}} \right)} \right)$, $\forall l$, which yields
 \begin{equation}\label{optimalVectorVMRT}
 {{\bf{v}}_l^{\star}} = {e^{j\left( {\arg \left( {{\alpha _l}{\rm{diag}}\left( {{\bf{h}}_l^H} \right){{\bf{a}}_R}\left( {{\phi _l}} \right)} \right) + \arg \left( {{\alpha _0}} \right)} \right)}}, \ \forall l \in \left[1,L\right].
 \end{equation}
 By substituting \eqref{optimalVectorVMRT} into \eqref{IRSMRTSNR}, the resulting SNR is
 \begin{equation}\label{IRSMRTSNR2}
 \begin{aligned}
 \gamma  &= \frac{{{N_t}{{\left( {\sqrt {{p_0}} \left| {{\alpha _0}} \right| + \sum\limits_{l = 1}^L {\sum\limits_{m = 1}^M {\sqrt {{p_l}} \left| {{\alpha _l}{h}_{l,m}^*{{a}_{R,m}}\left( {{\phi _l}} \right)} \right|} } } \right)}^2}}}{{{\sigma ^2}}}\\
 & = \frac{{{N_t}{{\left( {\sqrt {{p_0}} \left| {{\alpha _0}} \right| + \sum\limits_{l = 1}^L {\sqrt {{p_l}} \left| {{\alpha _l}} \right|{{\left\| {{{\bf{h}}_l}} \right\|}_1}} } \right)}^2}}}{{{\sigma ^2}}},
 \end{aligned}
 \end{equation}
 where ${h}_{l,m}^*$ and ${{a}_{R,m}}\left( {{\phi _l}} \right)$ are the $m$th element of ${{\bf{h}}_l^H}$ and ${{\bf{a}}_R}\left( {{\phi _l}} \right)$, respectively, and we have used the property that the receive array response vector ${{\bf{a}}_R}\left( {{\phi _l}} \right)$ has unit modulus for each element, i.e., $\left| {{a_{R,m}}\left( {{\phi _l}} \right)} \right| = 1$. By considering the transmit power constraint in \eqref{transmitPowerConstraint}, the optimal power allocation to maximize the SNR in \eqref{IRSMRTSNR2} can be obtained based on the Cauchy-Schwarz inequality, which is given by
 \begin{equation}\label{optimalPowerAllocationMRT}
 \left\{ \begin{split}
 &p_0^ \star  = \frac{{P{{\left| {{\alpha _0}} \right|}^2}}}{{{{\left| {{\alpha _0}} \right|}^2} + \sum\limits_{i = 1}^L {{{\left| {{\alpha _i}} \right|}^2}\left\| {{{\bf{h}}_i}} \right\|_1^2} }},\\
 &p_l^ \star  = \frac{{P{{\left| {{\alpha _l}} \right|}^2}\left\| {{{\bf{h}}_l}} \right\|_1^2}}{{{{\left| {{\alpha _0}} \right|}^2} + \sum\limits_{i = 1}^L {{{\left| {{\alpha _i}} \right|}^2}\left\| {{{\bf{h}}_i}} \right\|_1^2} }},\ \forall l \in \left[ {1,L} \right].
 \end{split} \right.
 \end{equation}
 The resulting SNR is
 \begin{equation}\label{optimalSNRMRT}
 \gamma  = \bar P{N_t}\left( {{{\left| {{\alpha _0}} \right|}^2} + \sum\limits_{l = 1}^L {{{\left| {{\alpha _l}} \right|}^2}\left\| {{{\bf{h}}_l}} \right\|_1^2} } \right),
 \end{equation}
 where ${\bar P} \triangleq P/{\sigma ^2}$.

 \subsection{Centralized Versus Distributed IRS Deployment}
 As a side result, the closed-form expression \eqref{optimalSNRMRT} is quite useful to compare the centralized versus distributed IRS deployment strategies, by fixing the total number of reflecting elements to $LM$. For ease of elaboration, we consider the LoS-dominating channels between each IRS and the user, for which ${{\bf{h}}_l}$ can be modelled as
 \begin{equation}\label{IRSReceiverLoSLink}
 \begin{aligned}
 {{\bf{h}}_l} = {\beta _l}&{\left[ {1, \cdots ,{e^{ - j2\pi \left( {{M_h} - 1} \right)\bar d\sin {\vartheta _{l,e}}\cos {\vartheta _{l,a}}}}} \right]^T}\\
 &\otimes {\left[ {1, \cdots ,{e^{ - j2\pi \left( {{M_v} - 1} \right)\bar d\cos {\vartheta _{l,e}}}}} \right]^T},
 \end{aligned}
 \end{equation}
 where ${\beta _l}$ denotes the complex-valued path gain, $\bar d$ denotes the inter-element spacing normalized by signal wavelength, and ${\vartheta _{l,e}}$ and ${\vartheta _{l,a}}$ denote the corresponding elevation and azimuth AoDs, respectively. In this case, $\left| {{h_{l,m}}} \right| = \left| {{\beta _l}} \right|$, $\forall m$, and ${\left\| {{{\bf{h}}_l}} \right\|_1} = M\left| {{\beta _l}} \right|$. Therefore, the SNR in \eqref{optimalSNRMRT} reduces to
 \begin{equation}\label{LoSOptimalSNRMRT}
 \begin{aligned}
 \gamma_{\rm distributed} &= \bar P{N_t}\Big( {{{\left| {{\alpha _0}} \right|}^2} + {M^2}\sum\limits_{l = 1}^L {{{\left| {{\alpha _l}{\beta _l}} \right|}^2}} } \Big)\\
 &\le \bar P{N_t}\left( {{{\left| {{\alpha _0}} \right|}^2} + L{M^2}{{\left| {{\alpha _{{l_{\max }}}}{\beta _{{l_{\max }}}}} \right|}^2}} \right),
 \end{aligned}
 \end{equation}
 where ${l_{\max }} = \arg \mathop {\max }\limits_{1 \le l \le L} {\left| {{\alpha _l}{\beta _l}} \right|^2}$. It is observed from \eqref{LoSOptimalSNRMRT} that for the distributed IRS deployment with a total of $LM$ reflecting elements, the received SNR scales with $L{M^2}$, rather than ${\left( {LM} \right)^2}$ as for centralized deployment \cite{wu2019intelligent}. It is worth remarking that although the above result is derived for DAM time-dispersive communications, it is also applicable for narrowband systems as considered in \cite{wu2019intelligent}. By substituting $M$ with $LM$ and $L$ with 1 in \eqref{LoSOptimalSNRMRT}, we get the SNR result for centralized IRS deployment, i.e., one single IRS with $LM$ reflecting elements, given by
 \begin{equation}\label{LoSOptimalSNRMRTCollocated}
 {\gamma _{{\rm{centralized}}}} = \bar P{N_t}\left( {{{\left| {{\alpha _0}} \right|}^2} + {L^2}{M^2}{{\left| {{\alpha _1}{\beta _1}} \right|}^2}} \right).
 \end{equation}

 It then follows from \eqref{LoSOptimalSNRMRT} and \eqref{LoSOptimalSNRMRTCollocated} that when the centralized IRS is placed at the strongest path, i.e., $l_{\max}=1$, we have $\gamma_{\rm distributed} \le {\gamma _{{\rm{centralized}}}}$. This is expected since for centralized IRS deployment, the BS only needs to concentrate its power towards one single direction, whereas it needs to split its power towards multiple IRSs for distributed IRS deployment. However, one should not make the simple conclusion that centralized IRS is always superior to distributed one, since the above result is obtained based on the asymptotic analysis and only single-user scenario is considered. For multi-user or multi-cell scenarios, it is quite likely that distributed IRS deployment is necessary due to its higher macro-diversity than centralized deployment.

 The above asymptotic analysis with ${N_t} \gg L$ shows the capability of DAM to transform the time-dispersive channel to ISI-free AWGN channel with the simple path-based MRT beamforming and delay pre-compensation, which provides an important insight for applying the low-complexity DAM to address the ISI issue in large antenna array systems. In the following, we study the DAM design for the general case with finite $N_t$ and $L$, when the orthogonal properties in \eqref{asymptoticalProperty}-\eqref{asymptoticalProperty3} do not hold. In this case, the three standard beamforming design approaches, namely ZF, MMSE, and MRT, are studied, on the per-path basis tailored for DAM transmission.

\section{Path-Based ZF Beamforming For ISI-Free \\ DAM Communication}\label{sectionPathBasedZFBeamforming}
 In this section, we study the path-based ZF beamforming for multi-IRS aided communication, so as to completely eliminate the ISI in \eqref{IRSDelayCompensationReceivedSignal}. Note that the assumption of LoS-dominating links for BS-IRS and IRS-user used in Section \ref{sectionSystemModel} is no longer needed in this section. With path-based ZF beamforming, $\left\{ {{\bf{f}}_{l'}} \right\}_{l' = 0}^L$ are designed so that all the ISI terms in \eqref{IRSDelayCompensationReceivedSignal} are eliminated, i.e.,
 \begin{equation}\label{IRSDAMZFCondition1}
 {\bf{h}}_0^H{{\bf{f}}_{l'}} = 0,\ \forall l' \in \left[ {1,L} \right],
 \end{equation}
 \begin{equation}\label{IRSDAMZFCondition2}
 {\bf{v}}_l^H{\rm{diag}}\left( {{\bf{h}}_l^H} \right){{\bf{G}}_l}{{\bf{f}}_{l'}} = 0,\ \forall l \ne l',\ l \in \left[1,L \right], \ l' \in \left[0,L \right].
 \end{equation}

 Let ${\bf{\tilde h}}_0^H \triangleq {\bf{h}}_0^H$, ${{\bf{\tilde h}}_l^H} \triangleq {\bf{v}}_l^H{\rm{diag}}\left( {{\bf{h}}_l^H} \right){{\bf{G}}_l}$, $\forall l \in \left[1,L\right]$, ${{\bf{H}}_0} \in {{\mathbb C}^{{N_t} \times L}} \triangleq [ {{{{\bf{\tilde h}}}_1}, \cdots ,{{{\bf{\tilde h}}}_L}}]$, and ${{\bf{H}}_{l'}} \in {{\mathbb C}^{{N_t} \times L}} \triangleq [ {{{\bf{\tilde h}}_0},{{{\bf{\tilde h}}}_1}, \cdots ,{{{\bf{\tilde h}}}_{l' - 1}},{{{\bf{\tilde h}}}_{l' + 1}}, \cdots ,{{{\bf{\tilde h}}}_L}} ]$, $\forall l' \in \left[1,L\right]$. Then the path-based ZF constraints in \eqref{IRSDAMZFCondition1} and \eqref{IRSDAMZFCondition2} can be compactly written as
 \begin{equation}\label{ZFConstraintMatrix}
 {\bf{H}}_{l'}^H{{\bf{f}}_{l'}} = {{\bf{0}}_{L \times 1}},\ \forall l' \in \left[0,L\right].
 \end{equation}
 The above ZF constraints are feasible as long as ${\bf{H}}_{l'}^H$ has a non-empty null-space, which is true almost surely when ${N_t} \ge L + 1$.

 With \eqref{IRSDAMZFCondition1} and \eqref{IRSDAMZFCondition2}, the received signal in \eqref{IRSDelayCompensationReceivedSignal} reduces to
 \begin{equation}\label{IRSDelayCompensationReceivedSignalZF}
 y\left[ n \right] = \left( {{\bf{h}}_0^H{{\bf{f}}_0} + \sum\limits_{l = 1}^L {{\bf{v}}_l^H{\rm{diag}}\left( {{\bf{h}}_l^H} \right){{\bf{G}}_l}{{\bf{f}}_l}} } \right)s\left[ {n - {n_{\max }}} \right] + z\left[ n \right],
 \end{equation}
 which, similar to the asymptotic result \eqref{IRSDelayCompensationReceivedSignalMRT2}, is simply the symbol sequence $s\left[ n \right]$ delayed by one single delay $n_{\max}$ with a multiplicative gain contributed by all the $L+1$ multi-paths. Thus, the original frequency-selective channel is transformed into an ISI-free AWGN channel.
 The received SNR in \eqref{IRSDelayCompensationReceivedSignalZF} can be expressed as
 \begin{equation}\label{IRSDAMReceivedSNR}
 \gamma  = \frac{{{{\left| {{\bf{h}}_0^H{{\bf{f}}_0} + \sum\limits_{l = 1}^L {{\bf{v}}_l^H{\rm{diag}}\left( {{\bf{h}}_l^H} \right){{\bf{G}}_l}{{\bf{f}}_l}} } \right|}^2}}}{{{\sigma ^2}}}.
 \end{equation}

 The received SNR in \eqref{IRSDAMReceivedSNR} can be maximized by jointly optimizing the path-based beamforming vectors $\left\{ {{\bf{f}}_{l'}} \right\}_{l' = 0}^L$ and the phase shift vectors $\left\{ {{{\bf{v }}_l}} \right\}_{l = 1}^L$. To tackle the non-convex unit-modulus constraints associated with the reflecting elements, i.e., $\left| {{v_{l,m}}} \right| = 1$, $\forall l,m$, we first consider the relaxed problem by temporarily considering the constraints $\left| {{v} _{l,m}} \right| \le 1$, $\forall l,m$. Then the problem for path-based ZF DAM beamforming can be formulated as
 \begin{equation}\label{originalProblemZF}
 \begin{aligned}
 \left( {\rm{P{\text -}ZF}} \right)\ \mathop {\max }\limits_{\left\{ {{\bf{f}}_{l'}} \right\}_{l' = 0}^L, \left\{ {{{\bf{v }}_l}} \right\}_{l = 1}^L} &\ \ \frac{1}{{{\sigma ^2}}}{\left| {{\bf{h}}_0^H{{\bf{f}}_0} + \sum\limits_{l = 1}^L {{\bf{v}}_l^H{\rm{diag}}\left( {{\bf{h}}_l^H} \right){{\bf{G}}_l}{{\bf{f}}_l}} } \right|^2}\\
 {\rm{s.t.}}&\ \ \sum\limits_{l' = 0}^L {{{\left\| {{{\bf{f}}_{l'}}} \right\|}^2}}  \le P,\\
 &\ \ {\bf{h}}_0^H{{\bf{f}}_{l'}} = 0,\ \forall l' \in \left[ {1,L} \right],\\
 &\ \ {\bf{v}}_l^H{\rm{diag}}\left( {{\bf{h}}_l^H} \right){{\bf{G}}_l}{{\bf{f}}_{l'}} = 0,\ \forall l \ne l',\\
 &\ \ \ \ \ \ \ \ \ \ \ \ \ \ \ l \in \left[1,L \right],\ l' \in \left[0,L \right],\\
 &\ \ \left| {{v} _{l,m}} \right| \le 1,\ \ \forall l,m. \nonumber
 \end{aligned}
 \end{equation}
 Problem (P-ZF) is difficult to be directly solved since it is non-convex and the optimization variables $\left\{ {{\bf{f}}_{l'}} \right\}_{l' = 0}^L$ and $\left\{ {{{\bf{v }}_l}} \right\}_{l = 1}^L$ are intricately coupled with each other both in the objective function and constraints. To overcome the above challenges, we propose an alternating optimization technique to solve (P-ZF), by iteratively optimizing the path-based beamfroming $\left\{ {{{\bf{f}}_{l'}}} \right\}_{l' = 0}^L$ and the phase shift vectors $\left\{{{\bf{v }}_l}\right\} _{l = 1}^L$.

\subsection{Path-Based ZF Beamforming Optimization}\label{SectionZFBeamformingVectorsOptimization}
 For any given IRS phase shift vectors $\left\{{{\bf{v }}_l}\right\} _{l = 1}^L$, the sub-problem of (P-ZF) for optimizing the beamforming vectors $\left\{ {{{\bf{f}}_{l'}}} \right\}_{l' = 0}^L$ can be written as
 \begin{equation}\label{subProblem1}
 \begin{aligned}
 \left( {\rm{P{\text -}ZF1}} \right)\ \ \mathop {\max }\limits_{\left\{ {{\bf{f}}_{l'}} \right\}_{l' = 0}^L} &\ \ \frac{1}{{{\sigma ^2}}}{\left| {\sum\limits_{l' = 0}^L {{\bf{\tilde h}}_{l'}^H{{\bf{f}}_{l'}}} } \right|^2}\\
 {\rm{s.t.}}&\ \sum\limits_{l' = 0}^L {{{\left\| {{{\bf{f}}_{l'}}} \right\|}^2}}  \le P,\\
 &\ {\bf{\tilde h}}_0^H{{\bf{f}}_{l'}} = 0,\ \forall l' \in \left[ {1,L} \right],\\
 &\ {\bf{\tilde h}}_l^H{{\bf{f}}_{l'}} = 0,\ \forall l \ne l',\ l \in \left[1,L \right], \ l' \in \left[0,L \right], \nonumber
 \end{aligned}
 \end{equation}
 where ${\bf{\tilde h}}_0^H$ and ${\bf{\tilde h}}_l^H$ are defined below \eqref{IRSDAMZFCondition2}.

 Let ${\bf{H}} \in {{\mathbb C}^{{N_t} \times \left( {L + 1} \right)}} \triangleq [ {{{\bf{\tilde h}}_0},{{{\bf{\tilde h}}}_1}, \cdots ,{{{\bf{\tilde h}}}_L}} ]$, and ${\bf{F}} \in {{\mathbb C}^{{N_t} \times \left( {L + 1} \right)}} \triangleq \left[ {{{\bf{f}}_0},{{\bf{f}}_1}, \cdots ,{{\bf{f}}_L}} \right]$. Without loss of generality, ${\bf{F}}$ can be decomposed as ${\bf{W}}{{\bf{\Lambda }}^{1/2}}$, where ${\bf{W}} \in {{\mathbb C}^{{N_t} \times \left( {L + 1} \right)}} \triangleq \left[ {{{\bf{w}}_0},{{\bf{w}}_1}, \cdots ,{{\bf{w}}_L}} \right]$ is designed to ensure the ZF constraints \eqref{IRSDAMZFCondition1} and \eqref{IRSDAMZFCondition2}, and ${\bf{\Lambda }} \triangleq {\rm{diag}}\left( {{\mu _0},{\mu _1}, \cdots ,{\mu _L}} \right)$ contains the power allocation coefficient $\mu_{l'}$ so that the total power constraint $\sum\nolimits_{l' = 0}^L {{{\left\| {{{\bf{f}}_{l'}}} \right\|}^2}}  = P$ is satisfied. In the following, we first determine ${\bf{W}}$ by noting that one effective beamforming design that satisfies the ZF constraints \eqref{IRSDAMZFCondition1} and \eqref{IRSDAMZFCondition2} is by ensuring ${{\bf{H}}^H}{\bf{W}} = {{\bf{I}}_{L + 1}}$. With ${N_t} \ge L + 1$, the matrix ${\bf{W}}$ can be obtained by taking the right pseudo inverse of ${{\bf{H}}^H}$, i.e., ${\bf{W}} = {({{\bf{H}}^H})^ + } = {\bf{H}}{({{\bf{H}}^H}{\bf{H}})^{ - 1}}$. Thus, ${{\bf{f}}_{l'}}$ can be expressed as ${{\bf{f}}_{l'}} = \sqrt {{\mu _{l'}}} {{\bf{w}}_{l'}}$, where ${\bf{w}}_{l'}$ is the $\left(l'+1\right)$th column of ${({{\bf{H}}^H})}^ +$, and the remaining task is to find the optimal power allocation coefficient $\mu_{l'}$ by solving the following optimization problem:
 \begin{equation}\label{subProblem2EquivalentProblem}
 \begin{aligned}
 \mathop {\max }\limits_{\left\{ {\mu _{l'}} \right\}_{l' = 0}^L} &\ \ \frac{1}{{{\sigma ^2}}}{\left| {\sum\limits_{l' = 0}^L {\sqrt {{\mu _{l'}}} {\bf{\tilde h}}_{l'}^H{{\bf{w}}_{l'}}} } \right|^2} = \frac{1}{{{\sigma ^2}}}{\left| {\sum\limits_{l' = 0}^L {\sqrt {{\mu _{l'}}} } } \right|^2}\\
 {\rm{s.t.}} &\ \ \sum\limits_{l' = 0}^L {{\mu _{l'}}{{\left\| {{{\bf{w}}_{l'}}} \right\|}^2}}  \le P,
 \end{aligned}
 \end{equation}
 where we have used the relationship ${\bf{\tilde h}}_{l'}^H{{\bf{w}}_{l' }} = 1$ based on the property of pseudo inverse, $\forall l' \in \left[0,L\right]$.

 \begin{theorem}\label{PathBasedZFBeamformingTheorem}
 The optimal solution to \eqref{subProblem2EquivalentProblem} is
 \begin{equation}\label{optimalPowerAllocationCoeffZF}
 \mu _{l'}^{\star} = \frac{P}{{\sum\limits_{i = 0}^L {\frac{1}{{{{\left\| {{{\bf{w}}_i}} \right\|}^2}}}} }}\frac{1}{{{{\left\| {{{\bf{w}}_{l'}}} \right\|}^4}}}, \ \forall l',
 \end{equation}
 and the optimal beamforming vectors for (P-ZF1) are
 \begin{equation}\label{optimalTransmitBeamforming}
 {\bf{f}}_{l'}^ \star = \frac{{\sqrt P {{\bf{w}}_{l'}}}}{{\sqrt {\sum\limits_{i = 0}^L {\frac{1}{{{\left\| {{{\bf{w}}_{i}}} \right\|}^2}}} } {{\left\| {{{\bf{w}}_{l'}}} \right\|}^2}}}, \ \forall l'.
 \end{equation}
 The resulting SNR is
 \begin{equation}\label{optimalTransmitBeamformingSNR}
 \gamma  = \bar P\sum\limits_{l' = 0}^L {\frac{1}{{{{\left\| {{{\bf{w}}_{l'}}} \right\|}^2}}}}.
 \end{equation}
 \end{theorem}

 \begin{IEEEproof}
 Please refer to Appendix~\ref{proofOfPathBasedZFBeamformingTheorem}.
 \end{IEEEproof}

\subsection{Phase Shifts Optimization}
 For any given beamforming vectors $\left\{ {{{\bf{f}}_{l'}}} \right\}_{l' = 0}^L$ in (P-ZF), the sub-problem of phase shift optimization is given by
 \begin{equation}\label{subProblem2}
 \begin{aligned}
 \left( {\rm{P{\text -}ZF2}} \right)\ \ \mathop {\max }\limits_{ \left\{ {{{\bf{v }}_l}} \right\}_{l = 1}^L} &\ \ \frac{1}{\sigma ^2}{\left| {{\bf{h}}_0^H{\bf{f}}_0 + \sum\limits_{l = 1}^L {{\bf{v}}_l^H{\rm{diag}}\left( {{\bf{h}}_l^H} \right){{\bf{G}}_l}{\bf{f}}_l} } \right|^2}\\
 {\rm{s.t.}} &\ \ {\left| {{\bf{v}}_l^H{\rm{diag}}\left( {{\bf{h}}_l^H} \right){{\bf{G}}_l}{{\bf{f}}_{l'}}} \right|^2} \le 0,\ \forall l \ne l',\\
 &\ \ \ \ \ \ \ \ \ \ \ \ \ \ \ \ \ \ l \in \left[1,L \right],\ l' \in \left[0,L \right],\\
 &\ \ \left| {{{v} _{l,m}}} \right| \le 1, \ \forall l,m. \nonumber
 \end{aligned}
 \end{equation}
 Note that in (P-ZF2), we have expressed the ZF constraint of (P-ZF) by its equivalent quadratic form.

 Let ${\bf{\tilde v}} \in {{\mathbb C}^{\left( {LM + 1} \right) \times 1}} \triangleq {\left[ {{\bf{v}}_1^T, \cdots ,{\bf{v}}_L^T,1} \right]^T}$, ${\bf{\tilde f}} \in {{\mathbb C}^{\left( {LM + 1} \right) \times 1}}$ $\triangleq {[{({\rm{diag}}\left( {{\bf{h}}_1^H} \right){{\bf{G}}_1}{{\bf{f}}_1})^T}, \cdots ,{({\rm{diag}}\left( {{\bf{h}}_L^H} \right){{\bf{G}}_L}{{\bf{f}}_L})^T},{\bf{h}}_0^H{{\bf{f}}_0}]^T}$. Then the objective function of (P-ZF2) can be compactly written as ${| {{{{\bf{\tilde v}}}^H}{\bf{\tilde f}}} |^2}/{\sigma ^2}$. Furthermore, for any $l \in \left[1,L\right]$, define the selection matrix ${{\bf{S}}_l} \in {{\mathbb C}^{M \times \left( {LM + 1} \right)}}$ so that ${{\bf{v}}_l} = {{\bf{S}}_l}{\bf{\tilde v}}$, which can be obtained by setting the $\left( {\left( {l - 1} \right)M + i} \right)$th element of the $i$th row as one, and the other elements as zero, $\forall i\in \left[ 1,M \right]$. Therefore, the first constraint of (P-ZF2) can be expressed as ${| {{{{\bf{\tilde v}}}^H}{{\bf{b}}_{l,l'}}} |^2} \le 0$, $\forall l \ne l'$, $l \in \left[ {1,L} \right]$, $l' \in \left[ {0,L} \right]$, where ${{\bf{b}}_{l,l'}} \in {{\mathbb C}^{\left( {LM + 1} \right) \times 1}} \triangleq {\bf{S}}_l^H{\rm{diag}}\left( {{\bf{h}}_l^H} \right){{\bf{G}}_l}{{\bf{f}}_{l'}}$. Then (P-ZF2) can be compactly written as
 \begin{equation}\label{subProblem1EquivalentProblem2}
 \begin{aligned}
 \mathop {\max }\limits_{ {\bf{\tilde v}}} &\ \ \frac{1}{\sigma ^2}{\left| {{{{\bf{\tilde v}}}^H}{\bf{\tilde f}}} \right|^2}\\
 {\rm{s.t.}} &\ \ {\left| {{{{\bf{\tilde v}}}^H}{{\bf{b}}_{l,l'}}} \right|^2} \le 0,\ \forall l \ne l',\ l \in \left[1,L \right], \ l' \in \left[0,L \right],\\
 &\ \ \left| {{{\tilde v}}_{m}} \right| \le 1, \ m \in \left[ {1,LM} \right],\\
 &\ \ {\tilde v_{LM + 1}} = 1,
 \end{aligned}
 \end{equation}
 where ${\tilde v}_{m}$ is the $m$th element of ${\bf{\tilde v}}$. The objective function of \eqref{subProblem1EquivalentProblem2} is convex, the maximization of which is a non-convex optimization problem. Fortunately, effective solution can be obtained by successive convex approximation (SCA) technique, which is an efficient iterative algorithm that successively updates the optimization variable at each iteration.
 Specifically, let ${\bf{\tilde v}}_r$ be the resulting optimization variable after the $r$th iteration. We have
 \begin{equation}\label{taylorApproximationPhaseShift}
 {\left| {{{{\bf{\tilde v}}}^H}{\bf{\tilde f}}} \right|^2} \ge \left| {{{{\bf{\tilde v}}}^H}{\bf{\tilde f}}} \right|_{{\rm{lb}}}^2 \triangleq {\left| {{\bf{\tilde v}}_r^H{\bf{\tilde f}}} \right|^2} + 2{\mathop{\rm Re}\nolimits} \left\{ {{{\left( {{\bf{\tilde v}} - {{{\bf{\tilde v}}}_r}} \right)}^H}{\bf{\tilde f}}{{{\bf{\tilde f}}}^H}{{{\bf{\tilde v}}}_r}} \right\}.
 \end{equation}
 As a result, problem \eqref{subProblem1EquivalentProblem2} is lower-bounded by the following problem for any given ${\bf{\tilde v}}_r$
 \begin{equation}\label{subProblem1EquivalentProblem3}
 \begin{aligned}
 \mathop {\max }\limits_{ {\bf{\tilde v}}} &\ \ \frac{1}{\sigma ^2}\left| {{{{\bf{\tilde v}}}^H}{\bf{\tilde f}}} \right|_{{\rm{lb}}}^2\\
 {\rm{s.t.}} &\ \ {\left| {{{{\bf{\tilde v}}}^H}{{\bf{b}}_{l,l'}}} \right|^2} \le 0,\ \forall l \ne l',\ l \in \left[1,L \right], \ l' \in \left[0,L \right],\\
 &\ \ \left| {{{\tilde v}}_{ m}} \right| \le 1, \  m \in \left[ {1,LM} \right],\\
 &\ \ {\tilde v_{LM + 1}} = 1,
 \end{aligned}
 \end{equation}
 which is a convex optimization problem and can be efficiently solved by standard convex optimization tools, such as CVX.

 \begin{algorithm}[t]
 \caption{Alternating Optimization for Problem (P-ZF)}
 \label{alg1}
 \begin{algorithmic}[1]
 \STATE Initialize $\{ {{\bf{v}}_{l}^0} \}_{l = 1}^{L}$ and let $k=0$.
 \REPEAT
 \STATE For given $\{ {{\bf{v}}_{l}^{k}} \}_{l = 1}^{L}$, obtain the optimal solution to problem (P-ZF1) based on \eqref{optimalTransmitBeamforming}, which is denoted as $\{ {\bf{f}}_{l'}^{k + 1}\} _{l' = 0}^{L}$.
 \STATE For given $\{ {\bf{f}}_{l'}^{k + 1}\} _{l' = 0}^{L}$,
 solve problem \eqref{subProblem1EquivalentProblem3} via SCA technique, and denote the solution as $\left\{ {{\bf{v}}_{l}^{k+1}} \right\}_{l = 1}^{L}$.
 \STATE Update $k=k+1$.
 \UNTIL the fractional increase of the objective value is below a threshold $ \epsilon > 0$.
 \end{algorithmic}
 \end{algorithm}

 The main procedure for solving (P-ZF) is summarized in Algorithm~\ref{alg1}, with its convergence shown below.

 \begin{proposition}\label{PathBasedZFBeamformingConvergence}
 The objective value of (P-ZF) obtained by Algorithm~\ref{alg1} is guaranteed to converge.
 \end{proposition}

 \begin{IEEEproof}
 Please refer to Appendix~\ref{proofOfPathBasedZFBeamformingConvergence}.
 \end{IEEEproof}

 With the obtained phase shift vector ${\bf{\tilde v}}^{\star}$ to (P-ZF), we extract the phases of ${\bf{\tilde v}}^{\star}$ as the IRS phase shifts, so that it satisfies the unit-modulus constraint, i.e.,
 \begin{equation}\label{ZFPhaseShiftFinal}
 \theta _{l,m}^ \star  = -\arg \left( {\tilde v_{\left( {l - 1} \right)M + m}^ \star } \right),\ \forall l,m.
 \end{equation}
 However, such IRS phase shift may render the ZF constraints \eqref{IRSDAMZFCondition1} and \eqref{IRSDAMZFCondition2} no longer valid. Fortunately, we only need to perform one additional path-based ZF beamforming as in \eqref{optimalTransmitBeamforming} to ensure the ZF constraints \eqref{IRSDAMZFCondition1} and \eqref{IRSDAMZFCondition2} are satisfied. The main complexity of obtaining path-based ZF beamforming vectors lies in the matrix pseudo-inverse, which is on the order of ${\cal O}(N_t(L + 1)^2)$. The phase shifts optimization has the complexity of ${\cal O}({(LM + 1)^3})$. Thus, the required complexity for the overall algorithm is ${\cal O}({I_1}(N_t(L + 1)^2 + {(LM + 1)^3}))$, with $I_1$ denoting the number of iterations.

 It is worth mentioning that the path-based ZF beamforming achieves the ISI-free AWGN channel, which is feasible almost surely as long as ${N_t} \ge L+1$. In the following, we consider the more general case when the ZF constraints in \eqref{IRSDAMZFCondition1} and \eqref{IRSDAMZFCondition2} are infeasible or undesirable. To this end, the optimal path-based MMSE beamforming and the low-complexity path-based MRT beamforming are respectively studied in the following two sections.

\section{Path-Based MMSE Beamforming for \\ DAM Communication}\label{sectionPathBasedMMSEBeamforming}
 In this section, we study the joint path-based beamforming and phase shifts design to maximize the received SINR for the general case with residual ISI.

 To derive the SINR in \eqref{IRSDelayCompensationReceivedSignal}, we first re-express it by properly grouping those interfering symbols with identical delay difference, since they correspond to identical symbols \cite{zeng2018multi,lu2022delay}. For the cascaded multi-path channel set ${\cal L} \triangleq \left\{ {l:l = 1, \cdots ,L} \right\}$, let ${{\cal L}_{l}} \triangleq {\cal L} \backslash {l}$ denote all other multi-paths excluding path $l$. Furthermore, let ${\Delta _{l',l}} \triangleq {n_{l'}} - {n_{l}}$ represent the \emph{delay difference} between path $l'$ and $l$ \cite{lu2022delay}, $\forall {l},{l'} \in \left[0,L\right]$. For $\forall l \ne l'$, we have ${{\Delta _{l',l}}} \ne 0$. Then $\forall l \ne l'$, ${\Delta _{l',l}} \in \left\{ { \pm 1, \cdots , \pm {n_{\rm{span}}}} \right\}$. The received signal in \eqref{IRSDelayCompensationReceivedSignal} can be equivalently written as
 \begin{equation}\label{IRSReceivedSignalDelayDifference}
 \begin{aligned}
 &y\left[ n \right]  = \left( {{\bf{h}}_0^H{{\bf{f}}_0} + \sum\limits_{l = 1}^L {{\bf{v}}_l^H{\rm{diag}}\left( {{\bf{h}}_l^H} \right){{\bf{G}}_l}{{\bf{f}}_l}} } \right)s\left[ {n - {n_{\max }}} \right] +  \\
 &\sum\limits_{l' = 1}^L {{\bf{h}}_0^H{{\bf{f}}_{l'}}s\left[ {n - {n_{\max }} + {\Delta _{l',0}}} \right]}  + \\
 &\sum\limits_{l = 1}^L {\sum\limits_{l' = 0,l' \ne l}^L {{\bf{v}}_l^H{\rm{diag}}\left( {{\bf{h}}_l^H} \right){{\bf{G}}_l}{{\bf{f}}_{l'}}s\left[ {n - {n_{\max }} + {\Delta _{l',l}}} \right]} }  + z\left[ n \right].
 \end{aligned}
 \end{equation}

 To group those interfering symbols with identical delay difference, for each delay difference $i \in \left\{ { \pm 1, \cdots , \pm {n_{\rm{span}}}} \right\}$, we define the following effective channel
 \begin{equation}\label{effectiveChannel}
 {\bf{g}}_{l'}^H\left[ i \right] {\rm =} \left\{ \begin{split}
 &{\bf{h}}_0^H,\ \ \ \ \ \ \ \ \ \ \ \ \ \ \ \ \ {\rm if}\  {n_{l'}} - {n_0} = i,\\
 &{{\bf{v}}_l^H{\rm{diag}}\left( {{\bf{h}}_l^H} \right){{\bf{G}}_l}},\ {\rm if}\ \exists l \in {{\cal L}_{{\cal L} \cap l'}},\ {\rm{s.t.}}\ {n_{l'}} - {n_l} = i,\\
 &{\bf{0}},\ \ \ \ \ \ \ \ \ \ \ \ \ \ \ \ \ \ \ \ {\rm otherwise}.
 \end{split} \right.
 \end{equation}

 \begin{figure}[!t]
 \centering
 \centerline{\includegraphics[width=3.2in,height=1in]{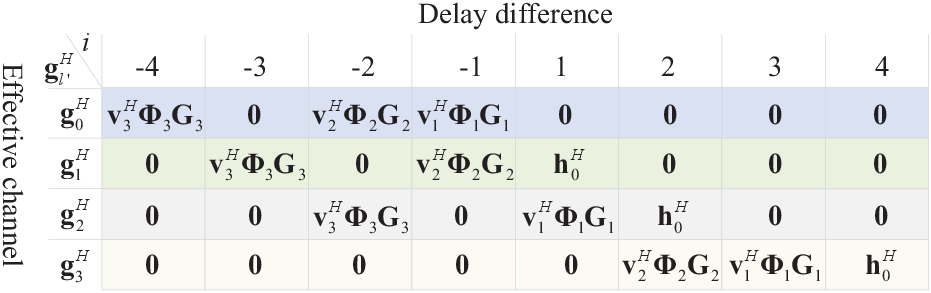}}
 \caption{An illustration of the effective channel with $L + 1 =4$ multi-paths, with $n_0 = 1$, $n_1 = 2$, $n_2 = 3$, and $n_3 = 5$. For the sake of brevity in the figure, ${{\bf{\Phi }}_l} \triangleq {\rm{diag}}\left( {{\bf{h}}_l^H} \right)$. The delay spread is ${n_{{\rm{span}}}} = {n_3} - {n_0} = 4$, and the delay difference $i \in \left\{ { \pm 1, \pm 2, \pm 3, \pm 4} \right\}$. For example, $n_1 - n_0 = 1$, ${\bf{g}}_1^H\left[ { 1} \right] = {\bf{h}}_0^H$, and $n_1 - n_3= -3$, ${\bf{g}}_1^H\left[ { - 3} \right] = {\bf{v}}_3^H{\rm{diag}}\left( {{\bf{h}}_3^H} \right){{\bf{G}}_3}$.}
 \label{delayDifferenceillustration}
 \end{figure}

 Fig.~\ref{delayDifferenceillustration} gives a simple illustration of the effective channel, with $L + 1= 4$ multi-paths. It is observed that for different multi-path pairs $\left( {l,l'} \right)$ in \eqref{IRSReceivedSignalDelayDifference}, their delay difference ${\Delta _{l',l}}$ might be identical, which implies that grouping their corresponding interfering symbols is necessary. As a result, \eqref{IRSReceivedSignalDelayDifference} can be equivalently written as
 \begin{equation}\label{IRSEquivalentReceivedSignalDelayDifference}
 \begin{aligned}
 y\left[ n \right] & = \left( {{\bf{h}}_0^H{{\bf{f}}_0} + \sum\limits_{l = 1}^L {{\bf{v}}_l^H{\rm{diag}}\left( {{\bf{h}}_l^H} \right){{\bf{G}}_l}{{\bf{f}}_l}} } \right)s\left[ {n - {n_{\max }}} \right] +  \\
 &\sum\limits_{i =  - {n_{{\rm{span}}}},i \ne 0}^{{n_{{\rm{span}}}}} {\left( {\sum\limits_{l' = 0}^L {{\bf{g}}_{l'}^H\left[ i \right]{{\bf{f}}_{l'}}} } \right)s} \left[ {n - {n_{\max }} + i} \right] + z\left[ n \right].
 \end{aligned}
 \end{equation}
 The resulting SINR is
 \begin{equation}\label{IRSEquivalentSINRDelayDifference}
 \gamma  = \frac{{{{\left| {{\bf{h}}_0^H{{\bf{f}}_0} + \sum\limits_{l = 1}^L {{\bf{v}}_l^H{\rm{diag}}\left( {{\bf{h}}_l^H} \right){{\bf{G}}_l}{{\bf{f}}_l}} } \right|}^2}}}{{\sum\limits_{i =  - {n_{{\rm{span}}}},i \ne 0}^{{n_{{\rm{span}}}}} {{{\left| {\sum\limits_{l' = 0}^L {{\bf{g}}_{l'}^H\left[ i \right]{{\bf{f}}_{l'}}} } \right|}^2}}  + {\sigma ^2}}}.
 \end{equation}

 With the SINR expression \eqref{IRSEquivalentSINRDelayDifference}, the optimization problem can be formulated as
 \begin{equation}\label{PathBasedMMSEProblem}
 \begin{aligned}
 \left( {\rm{P{\text -}MMSE}} \right) \mathop {\max }\limits_{\left\{ {{\bf{f}}_{l'}} \right\}_{l' = 0}^L, \left\{ {{{\bf{v }}_l}} \right\}_{l = 1}^L} &\ \frac{{{{\left| {{\bf{h}}_0^H{{\bf{f}}_0} + \sum\limits_{l = 1}^L {{\bf{v}}_l^H{\rm{diag}}\left( {{\bf{h}}_l^H} \right){{\bf{G}}_l}{{\bf{f}}_l}} } \right|}^2}}}{{\sum\limits_{i =  - {n_{{\rm{span}}}},i \ne 0}^{{n_{{\rm{span}}}}} {{{\left| {\sum\limits_{l' = 0}^L {{\bf{g}}_{l'}^H\left[ i \right]{{\bf{f}}_{l'}}} } \right|}^2}}  + {\sigma ^2}}}\\
 {\rm{s.t.}}&\ \ \sum\limits_{l' = 0}^L {{{\left\| {{{\bf{f}}_{l'}}} \right\|}^2}}  \le P,\\
 &\ \ \left| {{v} _{l,m}} \right| = 1, \ \forall l,m. \nonumber
 \end{aligned}
 \end{equation}
 Problem (P-MMSE) is difficult to be directly solved since the objective function consists of the effective channels defined in \eqref{effectiveChannel}, which are difficult to be explicitly expressed in terms of phase shifts. Besides, the optimization variables $\left\{ {{\bf{f}}_{l'}} \right\}_{l' = 0}^L$ and $\left\{ {{{\bf{v}}_l}} \right\}_{l = 1}^L$ are still intricately coupled with each other in the objective function. To tackle this problem, we propose an efficient solution based on the alternating optimization as in Section \ref{sectionPathBasedZFBeamforming}, where the path-based beamforming and phase shift vectors are optimized in an alternating manner, until the convergence is achieved.

 \subsection{Path-Based MMSE Beamforming Optimization}
 For any given IRS phase shift vectors $\left\{{{\bf{v }}_l}\right\} _{l = 1}^L$, the SINR in (P-MMSE) can be compactly written as
 \begin{equation}\label{IRSEquivalentSINRDelayDifferenceBeamforming}
 \gamma = \frac{{{{{\bf{\bar f}}}^H}{\bf{\bar h}}{{{\bf{\bar h}}}^H}{\bf{\bar f}}}}{{{{{\bf{\bar f}}}^H}\left( {\sum\limits_{i =  - {n_{{\rm{span}}}},i \ne 0}^{{n_{{\rm{span}}}}} {{\bf{\bar g}}\left[ i \right]{{{\bf{\bar g}}}^H}\left[ i \right]}  + {\sigma ^2}{\bf{I}}/{{\left\| {{\bf{\bar f}}} \right\|}^2}} \right){\bf{\bar f}}}},
 \end{equation}
 where ${\bf{\bar h}} \in {{\mathbb C}^{\left( {L + 1} \right){N_t} \times 1}} \triangleq [{\bf{h}}_0^T,{({\bf{G}}_1^H{\rm{diag}}({{\bf{h}}_1}){{\bf{v}}_1})^T}, \cdots ,$ ${({\bf{G}}_L^H{\rm{diag}}({{\bf{h}}_L}){{\bf{v}}_L})^T}{]^T}$, ${\bf{\bar f}}\in {{\mathbb{C}}^{\left( {L + 1} \right){N_t} \times 1}} \triangleq {\left[ {{\bf{f}}_0^T, \cdots ,{\bf{f}}_L^T} \right]^T}$, and ${\bf{\bar g}}\left[ i \right]\in {{\mathbb{C}}^{\left( {L + 1} \right){N_t} \times 1}} \triangleq {\left[ {{\bf{g}}_0^T\left[ i \right], \cdots ,{\bf{g}}_L^T\left[ i \right]} \right]^T} $. Therefore, (P-MMSE) reduces to
 \begin{equation}\label{PathBasedMMSEBeamforming}
 \begin{aligned}
 \mathop {\max }\limits_{\bf{\bar f}} &\ \ \frac{{{{{\bf{\bar f}}}^H}{\bf{\bar h}}{{{\bf{\bar h}}}^H}{\bf{\bar f}}}}{{{{{\bf{\bar f}}}^H}\left( {\sum\limits_{i =  - {n_{{\rm{span}}}},i \ne 0}^{{n_{{\rm{span}}}}} {{\bf{\bar g}}\left[ i \right]{{{\bf{\bar g}}}^H}\left[ i \right]}  + {\sigma ^2}{\bf{I}}/{{\left\| {{\bf{\bar f}}} \right\|}^2}} \right){\bf{\bar f}}}}\\
 {\rm{s.t.}}&\ \ {\left\| {{\bf{\bar f}}} \right\|^2} \le P.
 \end{aligned}
 \end{equation}

 The objective function of \eqref{PathBasedMMSEBeamforming} is a generalized Rayleigh quotient with respect to ${{\bf{\bar f}}}$, whose optimal solution is the MMSE design given by
 \begin{equation}\label{residualISIMMSEBeamforming}
 {{\bf{\bar f}}}^{{\rm{MMSE}}} = \sqrt P \frac{{{{\bf{C}}^{ - 1}}{\bf{\bar h}}}}{{\left\| {{{\bf{C}}^{ - 1}}{\bf{\bar h}}} \right\|}},
 \end{equation}
 where ${\bf{C}} \triangleq \sum\nolimits_{i =  - {n_{{\rm{span}}}},i \ne 0}^{{n_{{\rm{span}}}}} {{\bf{\bar g}}\left[ i \right]{{{\bf{\bar g}}}^H}\left[ i \right]}  + \frac{{{\sigma ^2}}}{P}{\bf{I}}$ denotes the interference-plus-noise covariance matrix. The resulting SINR is ${\gamma} = {{{\bf{\bar h}}}^H}{{\bf{C}}^{ - 1}}{\bf{\bar h}}$.

\subsection{Phase Shifts Optimization}
 For any given path-based beamforming vectors $\left\{ {{\bf{f}}_{l'}} \right\}_{l' = 0}^L$, the phases shift vectors are then optimized. Note that in (P-MMSE), the phase shift vectors $\left\{ {{{\bf{v}}_l}} \right\}_{l = 1}^L$ are implicitly included in ${\bf{g}}_{l'}^H\left[ i \right]$ in the denominator of the objective function, which is difficult to be directly optimized. As a result, we need to explicitly express the SINR in terms of the phase shift vectors.

 Let ${{\cal L}'} \triangleq \left\{ {l':l' = 0, \cdots ,L} \right\}$ be the set of all the multi-paths. In order to express the SINR of \eqref{IRSReceivedSignalDelayDifference} in terms of $\left\{{{\bf{v }}_l}\right\} _{l = 1}^L$, for each delay difference $i \in \left\{ { \pm 1, \cdots , \pm {n_{\rm{span}}}} \right\}$, define the following vector
 \begin{equation}
 {{\bf{e}}_0}\left[ i \right] = \left\{ \begin{split}
 &{{\bf{f}}_{l'}},\ {\rm if} \ \exists l' \in {\cal L},\ {\rm s.t.}\ {n_{l'}} - {n_0} = i,\\
 &{\bf{0}},\ \ {\rm otherwise},
 \end{split} \right.
 \end{equation}
 and $\forall l \in \left[ {1,L} \right]$,
 \begin{equation}
 {{\bf{e}}_l}\left[ i \right] = \left\{ \begin{split}
 &{{\bf{f}}_{l'}},\ {\rm if}\ \exists l' \in {{\cal L}'_l},\ {\rm s.t.} \ {n_{l'}} - {n_l} = i,\\
 &{\bf{0}},\ \ {\rm otherwise}.
 \end{split} \right.
 \end{equation}
 \begin{figure}[!t]
  \setlength{\abovecaptionskip}{-0.1cm}
 \setlength{\belowcaptionskip}{-0.1cm}
 \centering
 \centerline{\includegraphics[width=3.2in,height=1in]{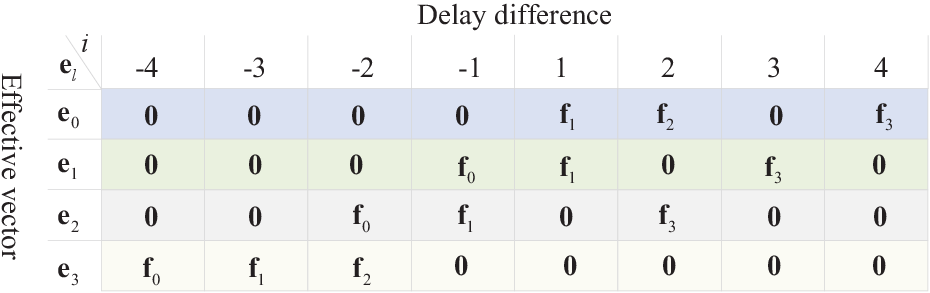}}
 \caption{An illustration of the effective vector with $L + 1 =4$ multi-paths, with $n_0 = 1$, $n_1 = 2$, $n_2 = 3$, and $n_3 = 5$. The delay spread is ${n_{{\rm{span}}}} = {n_3} - {n_0} = 4$, and the delay difference $i \in \left\{ { \pm 1, \pm 2, \pm 3, \pm 4} \right\}$. For example, $n_1 - n_0 = 1$, ${{\bf{e}}_0}\left[ 1 \right] = {{\bf{f}}_1}$, and $n_1 - n_3= -3$, ${{\bf{e}}_3}\left[ { - 3} \right] = {{\bf{f}}_1}$.}
 \label{delayDifferencePhaseillustration}
 \end{figure}

 Similar to Fig.~\ref{delayDifferenceillustration}, an illustration of ${{\bf{e}}_{l'}}\left[ i \right]$ with $L + 1= 4$ multi-paths is given in Fig.~\ref{delayDifferencePhaseillustration}. Then \eqref{IRSReceivedSignalDelayDifference} can be equivalently written as
 \begin{equation}\label{IRSReceivedSignalDelayDifferencePhase2}
 \begin{aligned}
 y\left[ n \right]  =& \left( {{\bf{h}}_0^H{{\bf{f}}_0} + \sum\limits_{l = 1}^L {{\bf{v}}_l^H{\rm{diag}}\left( {{\bf{h}}_l^H} \right){{\bf{G}}_l}{{\bf{f}}_l}} } \right)s\left[ {n - {n_{\max }}} \right] +  \\
 &\sum\limits_{i =  - {n_{{\rm{span}}}},i \ne 0}^{{n_{{\rm{span}}}}} {\Big( {\sum\limits_{l = 1}^L {{\bf{v}}_l^H{\rm{diag}}\left( {{\bf{h}}_l^H} \right){{\bf{G}}_l}{{\bf{e}}_l}\left[ i \right]} } \Bigg.} \\
 &\ \ \ \ \ \ \ \ \ \ \ \ \ \ \ \Big. { + {\bf{h}}_0^H{{\bf{e}}_0}\left[ i \right]} \Big)s\left[ {n - {n_{\max }} + i} \right] + z\left[ n \right].
 \end{aligned}
 \end{equation}
 The resulting SINR can be expressed as
 \begin{equation}\label{IRSEquivalentSINRDelayDifferencePhase}
 \hspace{-1ex}
 \begin{aligned}
 \gamma  &= \frac{{{{\left| {{\bf{h}}_0^H{{\bf{f}}_0} + \sum\limits_{l = 1}^L {{\bf{v}}_l^H{\rm{diag}}\left( {{\bf{h}}_l^H} \right){{\bf{G}}_l}{{\bf{f}}_l}} } \right|}^2}}}{{\sum\limits_{i =  - {n_{{\rm{span}}}},i \ne 0}^{{n_{{\rm{span}}}}} {{{\left| {\sum\limits_{l = 1}^L {{\bf{v}}_l^H{\rm{diag}}\left( {{\bf{h}}_l^H} \right){{\bf{G}}_l}{{\bf{e}}_l}\left[ i \right]}  +
 {\bf{h}}_0^H{{\bf{e}}_0}\left[ i \right]} \right|}^2}}  + {\sigma ^2}}}\\
 &= \frac{{{{\left| {{{{\bf{\tilde v}}}^H}{\bf{\tilde f}}} \right|}^2}}}{{\sum\limits_{i =  - {n_{{\rm{span}}}},i \ne 0}^{{n_{{\rm{span}}}}} {{{\left| {{{{\bf{\tilde v}}}^H}{\bf{\tilde e}}\left[ i \right]} \right|}^2}}  + {\sigma ^2}}}\\
 &=  \frac{{{{{\bf{\tilde v}}}^H}{\bf{\tilde f}}{{{\bf{\tilde f}}}^H}{\bf{\tilde v}}}}{{{{{\bf{\tilde v}}}^H}\left( {\sum\limits_{i =  - {n_{{\rm{span}}}},i \ne 0}^{{n_{{\rm{span}}}}} {{\bf{\tilde e}}\left[ i \right]{{{\bf{\tilde e}}}^H}\left[ i \right]}  + {\sigma ^2}{\bf{I}}/{{\left\| {{\bf{\tilde v}}} \right\|}^2}} \right){\bf{\tilde v}}}},
 \end{aligned}
 \end{equation}
 where ${\bf{\tilde e}}\left[ i \right]$ $ \in {{\mathbb C}^{\left( {LM + 1} \right) \times 1}} \triangleq [{({\rm{diag}}\left( {{\bf{h}}_1^H} \right){{\bf{G}}_1}{{\bf{e}}_1}\left[ i \right])^T}, \cdots ,$ ${({\rm{diag}}\left( {{\bf{h}}_L^H} \right){{\bf{G}}_L}{{\bf{e}}_L}\left[ i \right])^T},{\bf{h}}_0^H{{\bf{e}}_0}\left[ i \right]{]^T}$, and ${\bf{\tilde v}}$ and ${\bf{\tilde f}}$ are defined below (P-ZF2). Thus, for any given $\left\{ {{\bf{f}}_{l'}} \right\}_{l' = 0}^L$, (P-MMSE) is reduced to
 \begin{equation}\label{PathBasedMMSEPhaseShift}
 \begin{aligned}
 \mathop {\max }\limits_{\bf{\tilde v}} &\ \ \frac{{{{{\bf{\tilde v}}}^H}{\bf{\tilde f}}{{{\bf{\tilde f}}}^H}{\bf{\tilde v}}}}{{{{{\bf{\tilde v}}}^H}\left( {\sum\limits_{i =  - {n_{{\rm{span}}}},i \ne 0}^{{n_{{\rm{span}}}}} {{\bf{\tilde e}}\left[ i \right]{{{\bf{\tilde e}}}^H}\left[ i \right]}  + {\sigma ^2}{\bf{I}}/{{\left\| {{\bf{\tilde v}}} \right\|}^2}} \right){\bf{\tilde v}}}}\\
 {\rm{s.t.}}&\ \ \left| {{{\tilde v}}_{m}} \right| = 1, \ m \in \left[ {1,LM} \right],\\
 &\ \ {\tilde v_{LM + 1}} = 1.
 \end{aligned}
 \end{equation}

 By temporarily removing the constraints $\left| {{{\tilde v}}_{ m}} \right| = 1$, $ m \in \left[ {1,LM} \right]$, and ${\tilde v_{LM + 1}} = 1$, problem \eqref{PathBasedMMSEPhaseShift} is a generalized Rayleigh quotient with respect to $\bf{\tilde v}$, which is maximized by the MMSE solution, i.e.,
 \begin{equation}
 {{{\bf{\tilde v}}}^{\rm{MMSE}}} = \frac{{{{{\bf{\tilde C}}}^{ - 1}}{\bf{\tilde f}}}}{{\left\| {{{{\bf{\tilde C}}}^{ - 1}}{\bf{\tilde f}}} \right\|}},
 \end{equation}
 where ${\bf{\tilde C}} \triangleq \sum\nolimits_{i =  - {n_{{\rm{span}}}},i \ne 0}^{{n_{{\rm{span}}}}} {{\bf{\tilde e}}\left[ i \right]{{{\bf{\tilde e}}}^H}\left[ i \right]}  + {\sigma ^2}{\bf{I}}$. With the obtained solution ${{\bf{\tilde v}}}^{\rm{MMSE}}$, we extract the phases of ${{\bf{\tilde v}}}^{\rm{MMSE}}$ as the corresponding IRS phase shifts, so that it satisfies the constraints $\left| {{\tilde v}_{ m}} \right| = 1$, i.e., $\theta _{l,m}^{{\rm{MMSE}}} =  - \arg ( {\tilde v_{\left( {l - 1} \right)M + m}^{{\rm{MMSE}}}} )$, $\forall l,m$. Besides, the iteration procedure for the path-based MMSE beamforming scheme is similar to that of Algorithm~\ref{alg1}, which is omitted for brevity. The complexity of obtaining path-based MMSE beamforming vectors mainly lies in the matrix inversion, which requires ${\cal O}({((L + 1){N_t})^3})$. Similarly, the phase shifts optimization has the complexity of ${\cal O}({(LM + 1)^3})$. As a result, the overall complexity is given by ${\cal O}({I_2}({((L + 1){N_t})^3} + {(LM + 1)^3}))$, with $I_2$ denoting the number of iterations.

\section{Path-Based MRT Beamforming for \\DAM Communication}\label{sectionPathBasedMRTBeamforming}
 In this section, we study the path-based MRT beamforming for DAM transmission. Note that under the assumption of LoS-dominating channels between the BS and IRS, the path-based MRT beamforming is given in \eqref{pathBasedMRTBeamformingAsymptotical}, where each beamforming vector simply matches the associated transmit array response vector and is independent of the phase shift of each IRS. By discarding the LoS-dominating assumption, we study the MRT beamforming scheme based on the input-output relationship in \eqref{IRSDelayCompensationReceivedSignal}, where $\left\{ {{{\bf{f}}_{l'}}} \right\}_{l' = 0}^L$ and $\left\{{{\bf{v}}_l}\right\} _{l = 1}^L$ are designed to maximize the power of the desired signal, while ignoring the ISI terms. The problem is formulated as
 \begin{equation}\label{originalProblemMRT}
 \begin{aligned}
 \left( {\rm{P{\text -}MRT}} \right)\ \mathop {\max }\limits_{\left\{ {{{\bf{f}}_{l'}}} \right\}_{l' = 0}^L,\left\{ {{{\bf{v }}_l}} \right\}_{l = 1}^L} &\ \ {\left| {{\bf{h}}_0^H{{\bf{f}}_0} + \sum\limits_{l = 1}^L {{\bf{v}}_l^H{\rm{diag}}\left( {{\bf{h}}_l^H} \right){{\bf{G}}_l}{{\bf{f}}_l}} } \right|^2} \\
 {\rm{s.t.}}&\ \ \sum\limits_{l' = 0}^L {{{\left\| {{{\bf{f}}_{l'}}} \right\|}^2}}  \le P,\\
 &\ \ \left| {{v_{l,m}}} \right| = 1, \ \forall l,m. \nonumber
 \end{aligned}
 \end{equation}

 \begin{theorem}\label{PathBasedMRTBeamformingTheorem}
 For any given IRS phase shift vectors $\left\{{{\bf{v }}_l}\right\} _{l = 1}^L$, the optimal path-based transmit beamforming solution to problem (P-MRT) is
 \begin{equation}\label{pathBasedMRTBeamforming}
 \left\{ \begin{split}
 &{{\bf{f}}_{0}^{\rm{MRT}}} = \frac{{\sqrt P {{\bf{h}}_0}}}{{\sqrt {{{\left\| {{{\bf{h}}_0}} \right\|}^2} + \sum\limits_{i = 1}^L {{{\left\| {{\bf{G}}_i^H{\rm{diag}}\left( {{{\bf{h}}_i}} \right){{\bf{v}}_i}} \right\|}^2}} } }},\\
 &{{\bf{f}}_{l}^{\rm{MRT}}} = \frac{{\sqrt P {\bf{G}}_l^H{\rm{diag}}\left( {{{\bf{h}}_l}} \right){{\bf{v}}_l}}}{{\sqrt {{{\left\| {{{\bf{h}}_0}} \right\|}^2} + \sum\limits_{i = 1}^L {{{\left\| {{\bf{G}}_i^H{\rm{diag}}\left( {{{\bf{h}}_i}} \right){{\bf{v}}_i}} \right\|}^2}} } }},\ \forall l \in \left[ {1,L} \right],
 \end{split} \right.
 \end{equation}
 and the resulting objective function reduces to
 \begin{equation}\label{PMRTObjectiveFunction}
 P\left( {{{\left\| {{{\bf{h}}_0}} \right\|}^2} + \sum\limits_{l = 1}^L {{{\left\| {{\bf{G}}_l^H{\rm{diag}}\left( {{{\bf{h}}_l}} \right){{\bf{v}}_l}} \right\|}^2}} } \right).
 \end{equation}
 \end{theorem}

 \begin{IEEEproof}
 With the definition of ${\bf{\tilde h}}_0^H$ and ${\bf{\tilde h}}_l^H$ below \eqref{IRSDAMZFCondition2}, for any given $\left\{{{\bf{v}}_l}\right\} _{l = 1}^L$, (P-MRT) can be written as
 \begin{equation}\label{reducedProblemMRT}
 \begin{aligned}
 \mathop {\max }\limits_{\left\{ {{{\bf{f}}_{l'}}} \right\}_{l' = 0}^L} &\ \ {\Big| {\sum\limits_{l' = 0}^L {{\bf{\tilde h}}_{l'}^H{{\bf{f}}_{l'}}} } \Big|^2} \\
 {\rm{s.t.}}&\ \ \sum\limits_{l' = 0}^L {{{\left\| {{{\bf{f}}_{l'}}} \right\|}^2}}  \le P.
 \end{aligned}
 \end{equation}
 The optimal solution to \eqref{reducedProblemMRT} can be obtained based on the Cauchy-Schwarz inequality, which is similar to the proof of Theorem \ref{PathBasedZFBeamformingTheorem}, and thus omitted for brevity. By substituting \eqref{pathBasedMRTBeamforming} into the objective function of (P-MRT), \eqref{PMRTObjectiveFunction} can be obtained. The proof is thus completed.
 \end{IEEEproof}

 With Theorem \ref{PathBasedMRTBeamformingTheorem}, (P-MRT) reduces to maximizing the desired signal power via IRS phase shifts optimization, i.e.,
 \begin{equation}\label{MRTPhaseShiftsProblemOriginal}
 \begin{aligned}
 \mathop {\max }\limits_{\left\{ {{{\bf{v}}_l}} \right\}_{l = 1}^L} &\ \ P\left( {{{\left\| {{{\bf{h}}_0}} \right\|}^2} + \sum\limits_{l = 1}^L {{{\left\| {{\bf{G}}_l^H{\rm{diag}}\left( {{{\bf{h}}_l}} \right){{\bf{v}}_l}} \right\|}^2}} } \right) \\
 {\rm{s.t.}}&\ \ \left| {{{v} _{l,m}}} \right| = 1, \ \forall l,m.
 \end{aligned}
 \end{equation}
 By discarding the constant term $P{\left\| {{{\bf{h}}_0}} \right\|^2}$, problem \eqref{MRTPhaseShiftsProblemOriginal} can be decomposed into $L$ parallel subproblems, each for one IRS. Specifically, for IRS $l$, the problem is
 \begin{equation}\label{MRTPhaseShiftsProblem}
 \begin{aligned}
 \mathop {\max }\limits_{{{\bf{v}}_l}}&\ \ {\left\| {{\bf{G}}_l^H{\rm{diag}}\left( {{{\bf{h}}_l}} \right){{\bf{v}}_l}} \right\|^2}\\
 {\rm{s.t.}}&\ \ \left| {{{{v}}_{l,m}}} \right| = 1,\ \forall m.
 \end{aligned}
 \end{equation}

 To tackle this non-convex problem, we propose an efficient iterative algorithm, where the contribution of each individual element ${v_{l,m}}$ to the objective function is extracted, with the remaining elements being fixed, and the phase shift of each element is updated alternately via the classic coordinate descent method \cite{sohrabi2016hybrid,huang2020achievable}. Specifically, let ${\bf{G}}_l^H{\rm{diag}}\left( {{{\bf{h}}_l}} \right) = \left[ {{{\bf{r}}_{l,1}}, \cdots ,{{\bf{r}}_{l,m}}, \cdots ,{{\bf{r}}_{l,M}}} \right]$, where ${{\bf{r}}_{l,m}} \in {{\mathbb C}^{{N_t} \times 1}}$ denotes the $m$th column of the matrix ${\bf{G}}_l^H{\rm{diag}}\left( {{{\bf{h}}_l}} \right)$. Then, by focusing on the $m$th reflecting element, the objective of problem \eqref{MRTPhaseShiftsProblem} can be expressed as
 \begin{equation}\label{MRTPhaseShiftsSubProblem}
 \begin{aligned}
 &{\left\| {{\bf{G}}_l^H{\rm{diag}}\left( {{{\bf{h}}_l}} \right){{\bf{v}}_l}} \right\|^2} = {\Big\| {\sum\limits_{m' \ne m}^{M} {{{\bf{r}}_{l,m'}}} {e^{ - j{\theta _{l,m'}}}} + {{\bf{r}}_{l,m}}{e^{ - j{\theta _{l,m}}}}} \Big\|^2}\\
 &= {\left\| {{{\bf{q}}_{l,m}}} \right\|^2} + {\left\| {{{\bf{r}}_{l,m}}} \right\|^2} + 2{\mathop{\rm Re}\nolimits} \left\{ {{e^{ - j{\theta _{l,m}}}}{\bf{q}}_{l,m}^H{{\bf{r}}_{l,m}}} \right\},
 \end{aligned}
 \end{equation}
 where ${{\bf{q}}_{l,m}} \triangleq \sum\nolimits_{m' \ne m}^{M} {{{\bf{r}}_{l,m'}}} {e^{ - j{\theta _{l,m'}}}}$. When the phase shifts of all IRS elements are fixed except that of the $m$th element ${v_{l,m}}$, the optimal ${\theta _{l,m}}$ to maximize \eqref{MRTPhaseShiftsSubProblem} is  \cite{sohrabi2016hybrid,huang2020achievable}
 \begin{equation}\label{MRTSinglePhaseShift}
 \theta _{l,m}^{\star} = \arg ({\bf{q}}_{l,m}^H{{\bf{r}}_{l,m}}).
 \end{equation}
 Therefore, we propose an iterative algorithm to sequentially update each element of ${{\bf{v}}_l}$ according to \eqref{MRTSinglePhaseShift}, which is summarized in Algorithm~\ref{alg2}. Note that since the objective function is non-decreasing in each iteration, the algorithm is guaranteed to converge.

 \begin{algorithm}[t]
 \caption{Iterative Algorithm for \eqref{MRTPhaseShiftsProblem}}
 \label{alg2}
 \begin{algorithmic}[1]
 \STATE Initialize $\{ {{\bf{v}}_{l}} \}_{l = 1}^{L}$ and let $k=0$.
 \REPEAT
 \STATE \textbf{for} $m = 1:M$ \textbf{do}
 \STATE \quad Calculate ${{\bf{q}}_{l,m}}$ and obtain the optimal phase shift of \\
 \quad the $m$th reflecting element based on \eqref{MRTSinglePhaseShift}.
 \STATE \quad Obtain the objective function value of \eqref{MRTPhaseShiftsProblem}.
 \STATE \quad Update $k=k+1$.
 \STATE \textbf{end for}
 \UNTIL the fractional increase of the objective function of \eqref{MRTPhaseShiftsProblem} is below a threshold $ \epsilon > 0$.
 \end{algorithmic}
 \end{algorithm}

 By applying the path-based MRT beamforming \eqref{pathBasedMRTBeamforming} and the IRS phase shifts from Algorithm~\ref{alg2} to \eqref{IRSEquivalentSINRDelayDifference}, the SINR for MRT design can be obtained. The calculation of path-based MRT beamforming vectors has the complexity of ${\cal O}(L{N_t}{M^2})$. The complexity for optimizing phase shifts is ${\cal O}({I_3}LM{N_t})$, with $I_3$ denoting the number of iterations for each reflecting element. Thus, the total complexity is given by ${\cal O}(L{N_t}{M^2} + {I_3}LM{N_t})$.

\section{Benchmark: OFDM-Based Multi-IRS\\ Aided Communication}\label{sectionBenchmarkOFDMScheme}
 In this subsection, we consider the benchmarking scheme of OFDM-based multi-IRS aided communication. Let $B$ and $K$ denote the total bandwidth and the number of OFDM sub-carriers, respectively, and ${{\bf{u}}_k} \in {{\mathbb C}^{{N_t} \times 1}}$ denote the frequency-domain beamforming vector for the $k$th sub-carrier. Denote by $T_c$ the channel coherence time, and $n_c = {T_c}/{T_s}$ the number of signal samples within each channel coherence time, with ${T_s} = 1/B$. Each OFDM symbol has duration $\left( {K + {N_{{\rm{CP}}}}} \right){T_s}$, and the number of OFDM symbols for each channel coherence time is ${n_{{\rm{OFDM}}}} = {n_c}/\left( {K + {N_{{\rm{CP}}}}} \right)$, where $N_{\rm CP}$ is the length of the CP.

 With the channel impulse response given in \eqref{discreteTimeChannelImpulseResponse}, the channel frequency response of the $k$th sub-carrier is obtained by applying $K$-point DFT, given by
 \begin{equation}\label{frequencyDomainChannelk}
 {{\bf{h}}^H}\left[ k \right] = \frac{1}{{\sqrt K }}\Big( {{\bf{h}}_0^H{e^{ - j\frac{{2\pi k{n_0}}}{K}}} + \sum\limits_{l = 1}^L {{\bf{v}}_l^H{\rm{diag}}\left( {{\bf{h}}_l^H} \right){{\bf{G}}_l}{e^{ - j\frac{{2\pi k{n_l}}}{K}}}} } \Big).
 \end{equation}
 The effective spectral efficiency by taking into account the CP overhead in bits/second/Hz (bps/Hz) is then given by
 \begin{equation}\label{effectiveSpectralEfficiencyOrigin}
 \begin{aligned}
 R_{\rm OFDM} &= \frac{{{n_c} - {n_{{\rm{OFDM}}}}{N_{{\rm{CP}}}}}}{{{n_c}}}\frac{1}{K}\sum\limits_{k = 0}^{K - 1} {{{\log }_2}\left( {1 + \frac{{{\left| {{{\bf{h}}^H}\left[ k \right]{{\bf{u}}_k}} \right|}^2}}{{{\sigma ^2}/K}}} \right)} \\
 &= \frac{1}{{K + {N_{{\rm{CP}}}}}}\sum\limits_{k = 0}^{K - 1} {{{\log }_2}\left( {1 + \frac{{{\left| {{{\bf{h}}^H}\left[ k \right]{{\bf{u}}_k}} \right|}^2}}{{{\sigma ^2}/K}}} \right)}.
 \end{aligned}
 \end{equation}
 It is obvious that for each subcarrier, the MRT beamforming is optimal for \eqref{effectiveSpectralEfficiencyOrigin}, i.e., ${{\bf{u}}_k} = \sqrt {{p_k}} {\bf{h}}\left[ k \right]/\left\| {{\bf{h}}\left[ k \right]} \right\|$, where $p_k$ is the transmit power of the $k$th sub-carrier. Thus, \eqref{effectiveSpectralEfficiencyOrigin} reduces to
 \begin{equation}\label{effectiveSpectralEfficiency1}
 {R_{{\rm{OFDM}}}} = \frac{1}{{K + {N_{{\rm{CP}}}}}}\sum\limits_{k = 0}^{K - 1} {{{\log }_2}\left( {1 + \frac{{K{p_k}{{\left\| {{\bf{h}}\left[ k \right]} \right\|}^2}}}{{{\sigma ^2}}}} \right)}.
 \end{equation}

 In the following, we temporarily assume that the total power $KP$ is equally allocated over $K$ sub-carriers, i.e., ${p_k} = P$, while leaving the power allocation problem after the BS beamforming and IRS phase shifts have been determined. By letting ${\bf{\tilde G}}[k] \in {\mathbb C}^{{{N_t} \times \left( {LM + 1} \right)}} \triangleq [{\bf{G}}_1^H{\rm diag}({{\bf{h}}_1}){e^{j\frac{{2\pi k{n_1}}}{K}}},$ $ \cdots ,{\bf{G}}_L^H{\rm diag}({{\bf{h}}_L}){e^{j\frac{{2\pi k{n_L}}}{K}}},{{\bf{h}}_0}{e^{j\frac{{2\pi k{n_0}}}{K}}}]$, the effective spectral efficiency of OFDM under the equal power allocation can be expressed as
 \begin{equation}\label{SEOFDMEqualPower}
 {R_{\rm OFDM}} = \frac{1}{{K + {N_{\rm CP}}}}\sum\limits_{k = 0}^{K - 1} {{{\log }_2}\left( {1 + \bar P{{\left\| {{\bf{\tilde G}}\left[ k \right]{\bf{\tilde v}}} \right\|}^2}} \right)},
 \end{equation}
 where $\bf{\tilde v}$ is defined after (P-ZF2). The phase shift vector $\bf{\tilde v}$ is then optimized to maximize the effective spectral efficiency in \eqref{SEOFDMEqualPower}. By dropping the constant term in ${R_{\rm OFDM}}$, the problem can be formulated as
 \begin{equation}\label{OFDMSEProblem}
 \begin{aligned}
 \left( {\rm{P{\text -}OFDM}} \right)\ \ \mathop {\max }\limits_{\bf{\tilde v}}&\ \ \sum\limits_{k = 0}^{K - 1} {{{\log }_2}\left( {1 + \bar P{{\left\| {{\bf{\tilde G}}\left[ k \right]{\bf{\tilde v}}} \right\|}^2}} \right)} \\
 {\rm{s.t.}}&\ \ \left| {{{\tilde v}}_{ m}} \right| \le 1, \  m \in \left[ {1,LM} \right],\\
 &\ \ {\tilde v_{LM + 1}} = 1, \nonumber
 \end{aligned}
 \end{equation}
 where the non-convex unit-modulus constraints associated with the reflecting elements are temporarily relaxed. Similar to problem \eqref{subProblem1EquivalentProblem2}, (P-OFDM) can also be solved via SCA technique. Specifically, for any given local point ${\bf{\tilde v}}_r$ at the $r$th iteration, we have
 \begin{equation}\label{OFDMConvexTermLowerBound}
 \begin{aligned}
 {\left\| {{\bf{\tilde G}}\left[ k \right]{\bf{\tilde v}}} \right\|^2}& \ge \left\| {{\bf{\tilde G}}\left[ k \right]{\bf{\tilde v}}} \right\|_{{\rm{lb}}}^2 \triangleq {\left\| {{\bf{\tilde G}}\left[ k \right]{{{\bf{\tilde v}}}_r}} \right\|^2} + \\
 &\ \ \ \  2{\mathop{\rm Re}\nolimits} \left\{ {{{\left( {{\bf{\tilde v}} - {{{\bf{\tilde v}}}_r}} \right)}^H}{{{\bf{\tilde G}}}^H}\left[ k \right]{\bf{\tilde G}}\left[ k \right]{{{\bf{\tilde v}}}_r}} \right\}.
 \end{aligned}
 \end{equation}
 By replacing the objective function of (P-OFDM) with its lower bound, we obtain the following problem
 \begin{equation}\label{OFDMSELowerBoundProblem}
 \begin{aligned}
 \mathop {\max }\limits_{\bf{\tilde v}}&\ \ \sum\limits_{k = 0}^{K - 1} {{{\log }_2}\left( {1 + \bar P{{\left\| {{\bf{\tilde G}}\left[ k \right]{\bf{\tilde v}}} \right\|}_{\rm lb}^2}} \right)} \\
 {\rm{s.t.}}&\ \ \left| {{{\tilde v}}_{ m}} \right| \le 1, \  m \in \left[ {1,LM} \right],\\
 &\ \ {\tilde v_{LM + 1}} = 1,
 \end{aligned}
 \end{equation}
 which is a convex optimization problem and can be solved CVX. Thus, (P-OFDM) can be solved by iteratively optimizing \eqref{OFDMSELowerBoundProblem} with the local point $\bf{\tilde v}$ updated in each iteration. With the obtained phase shift vector, its phases are then extracted as the IRS phase shifts as in \eqref{ZFPhaseShiftFinal}, so that the unit-modulus constraint is satisfied. Furthermore, by substituting the achieved IRS phase shift into \eqref{effectiveSpectralEfficiency1}, the power allocation can be obtained by the classic water-filling (WF) algorithm.

 Last, we study the impact of guard interval overhead for OFDM and ZF, MMSE, and MRT DAM. The CP length for OFDM is chosen as ${N_{\rm{CP}}} = {n_{\max }}$, and the CP overhead is thus ${n_{{\rm{OFDM}}}}{n_{\max }}/{n_c} = {n_{\max }}/\left( {{n_{\max }} + K} \right)$. By contrast, it is observed from \eqref{IRSDelayCompensationReceivedSignal} that the received signal for DAM is at most delayed by ${n_{\max }} + {n_{{\rm{span}}}} \approx 2{n_{\max }}$. To avoid the ISI across different channel coherence blocks, a guard interval of length $2{n_{\max }}$ is required for each coherence block. As a result, the effective spectral efficiency of ZF, MMSE, and MRT DAM is
 \begin{equation}
 {R_{{\rm{DAM}}}} = \frac{{{n_c} - 2{n_{\max }}}}{{{n_c}}}{\log _2}\left( {1 + {\gamma _a}} \right),
 \end{equation}
 where $a \in \left\{ {{\rm{ZF}},{\rm{MMSE}},{\rm{MRT}}} \right\}$, and ${{\gamma _a}}$ can be obtained based on Sections \ref{sectionPathBasedZFBeamforming}, \ref{sectionPathBasedMMSEBeamforming}, and \ref{sectionPathBasedMRTBeamforming}, respectively. It is observed that the guard interval overhead of DAM is $2{n_{\max }}/{n_c}$, which is significantly reduced than OFDM when ${n_{{\rm{OFDM}}}} \gg 2$.

\section{Simulation Results}\label{sectionNumericalResults}
\begin{figure}[!t]
 \centering
 \centerline{\includegraphics[width=3.0in,height=1.75in]{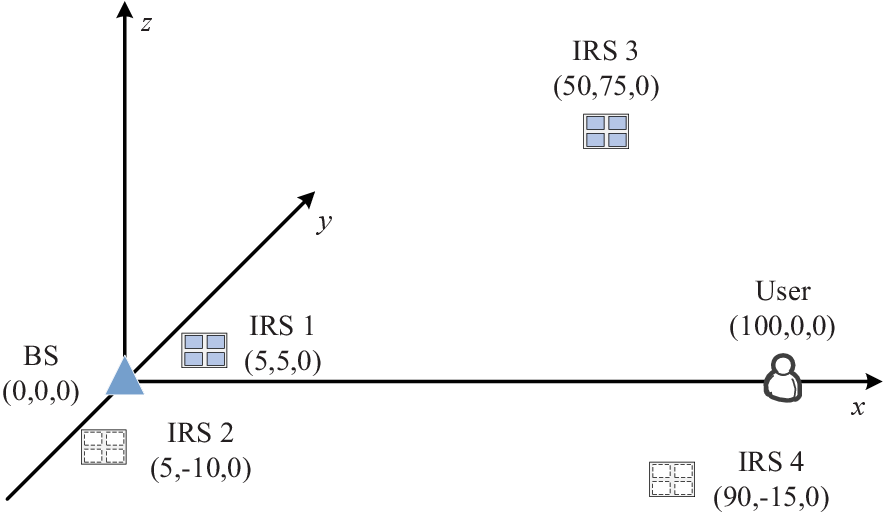}}
 \caption{Simulation setup of the multi-IRS aided communication system.}
 \label{locationIllustration}
 \end{figure}

 In this section, simulation results are provided to evaluate the performance of DAM for multi-IRS aided communications. We consider a mmWave system at $28$ GHz, with its total bandwidth and the noise power spectrum density given by $B=128$ MHz and $N_0 = -174$ dBm/Hz, respectively. The channel coherence time is set as $T_c = 1$ ms. Thus, the total number of signal samples within each coherence time is ${n_c} = B{T_c} = 1.28\times{10^5}$. Under a three-dimensional (3D) Cartesian coordinate system, the BS and the user are located at $\left( {0,0,0} \right)$ and $\left( {100~{\rm m},0,0} \right)$, respectively. The number of IRSs is $L = 4$, whose locations are $\left( {5~{\rm m},5~{\rm m},0} \right)$, $\left( {5~{\rm m},-10~{\rm m},0} \right)$, $\left( {50~{\rm m},75~{\rm m},0} \right)$ and $\left( {90~{\rm m},-15~{\rm m},0} \right)$, respectively, as shown in Fig.~\ref{locationIllustration}. The discrete-time delays are then given by $n_0 = 43$, $n_1 =44$, $n_2 = 46$, $n_3=77$, and $n_4 = 47$, respectively. The upper bound of the maximum delay over all channel coherence blocks is ${\tilde n_{\max }} = 77$.
 Besides, the distance-dependent path loss model is given by \cite{wu2019intelligent}
 \begin{equation}
 L\left( d \right) = {C_0}{\left( {\frac{d}{{{D_0}}}} \right)^{ - \upsilon }},
 \end{equation}
 where $C_0$ is the path loss at the reference distance $D_0 = 1$ m, $d$ denotes the link distance, and $\upsilon$ is the path loss exponent. Specifically, the path loss exponent of the direct BS-user link is set as ${\upsilon _{{\rm{TU}}}} = 3.5$, and those of the BS-IRS and the IRS-user links are set as ${\upsilon _{{\rm{TI}}}} = {\upsilon _{{\rm{IU}}}} = 2$. For the direct channel from the BS to the user, the channel tap ${{\bf{h}}_0}$ is modelled by Rician fading with Rician factor $\zeta$. Unless otherwise stated, we set $\zeta  = 5$ dB. For the benchmarking scheme of OFDM-based multi-IRS aided communication, we set the number of sub-carriers as $K=512$, and the CP length is ${\tilde n_{\max }} = 77$. Thus, for each channel coherence time, the number of OFDM symbols is ${n_{{\rm{OFDM}}}} = {n_c}/\left( {K + {{\tilde n}_{\max }}} \right) \approx 217$.

 \begin{figure}[!t]
 \centering
 \centerline{\includegraphics[width=3.5in,height=2.625in]{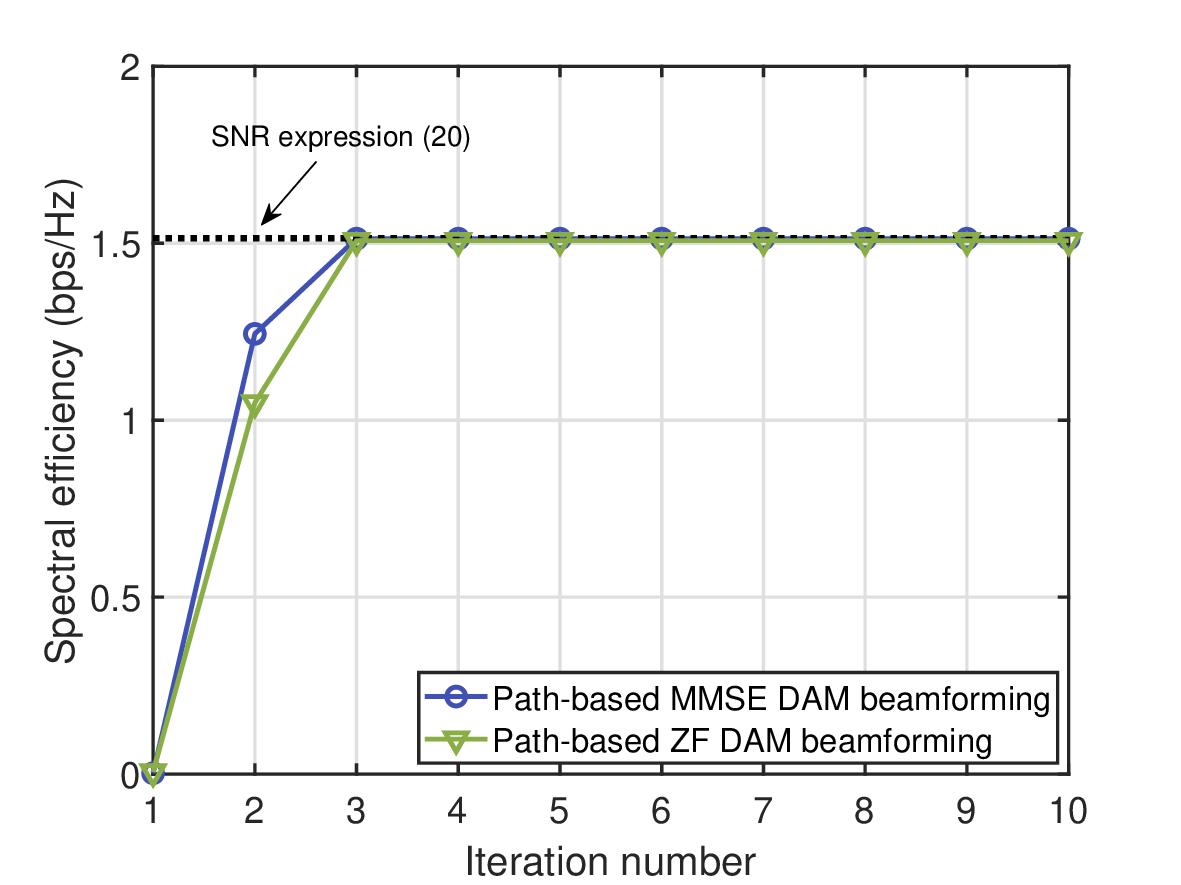}}
 \caption{Convergence plot of path-based MMSE and ZF DAM beamforming.}
 \label{pathBasedBeamformingConvergencePerformance}
 \end{figure}
\begin{figure}[!t]
 \centering
 \centerline{\includegraphics[width=3.5in,height=2.625in]{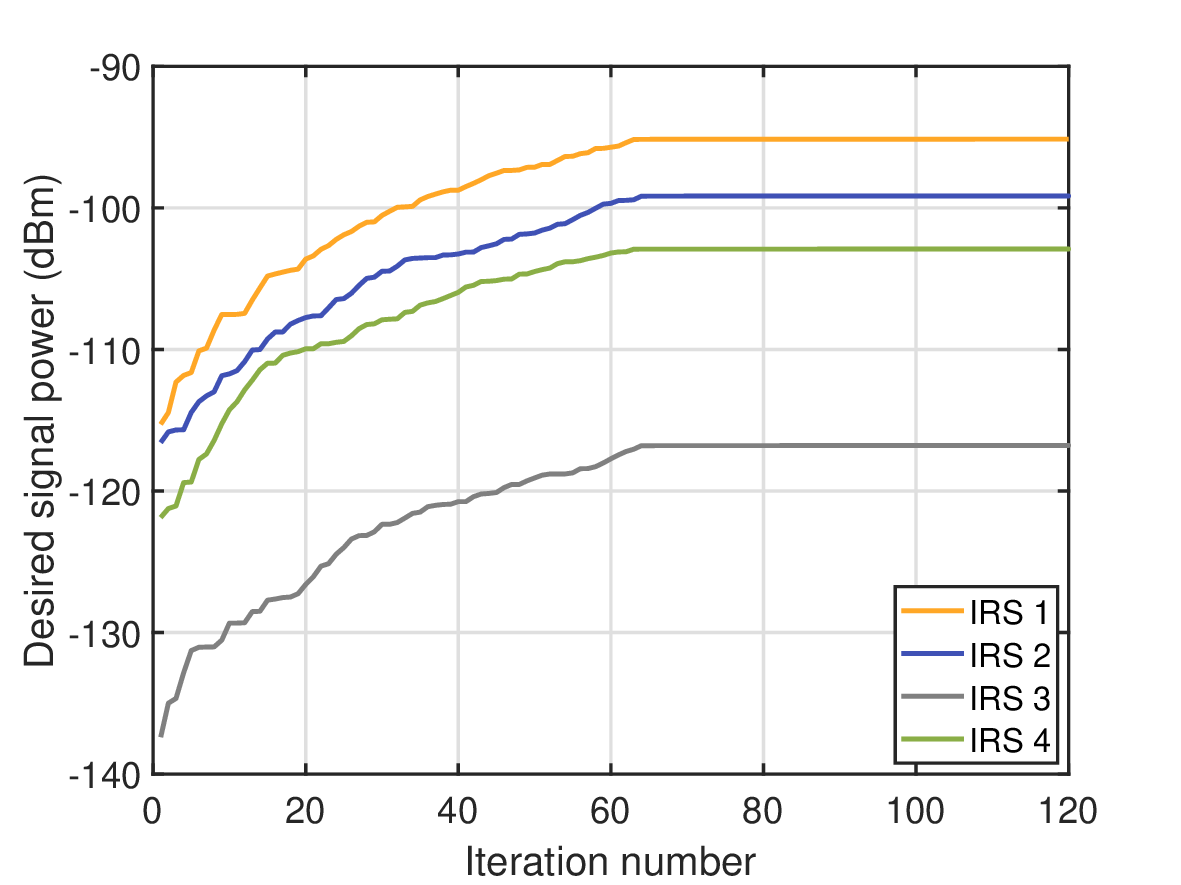}}
 \caption{Convergence plot of Algorithm~\ref{alg2}. }
 \label{PathBasedMRTconvergencePerformance}
 \end{figure}

 Fig.~\ref{pathBasedBeamformingConvergencePerformance} shows the convergence plot of path-based MMSE and ZF DAM beamforming for LoS-dominating channels. The number of antennas is ${N_t} = 64$, and that of reflecting elements for each IRS is $M = {M_h}{M_v} = 64$, with ${M_h} = {M_v} = 8$. The transmit power is $P =30$ dBm. It is observed that the two proposed DAM beamforming schemes converge quickly within several iterations. Furthermore, Fig.~\ref{PathBasedMRTconvergencePerformance} plots the convergence behaviour of the proposed iterative algorithm, i.e., Algorithm~\ref{alg2}, where one iteration refers to one update of each reflecting element phase shift. It is observed that the algorithm leads to a non-decreasing received signal power at the user over iterations from all IRSs, and eventually approaches a converged solution.


 \begin{figure}[!t]
 \centering
 \centerline{\includegraphics[width=3.5in,height=2.625in]{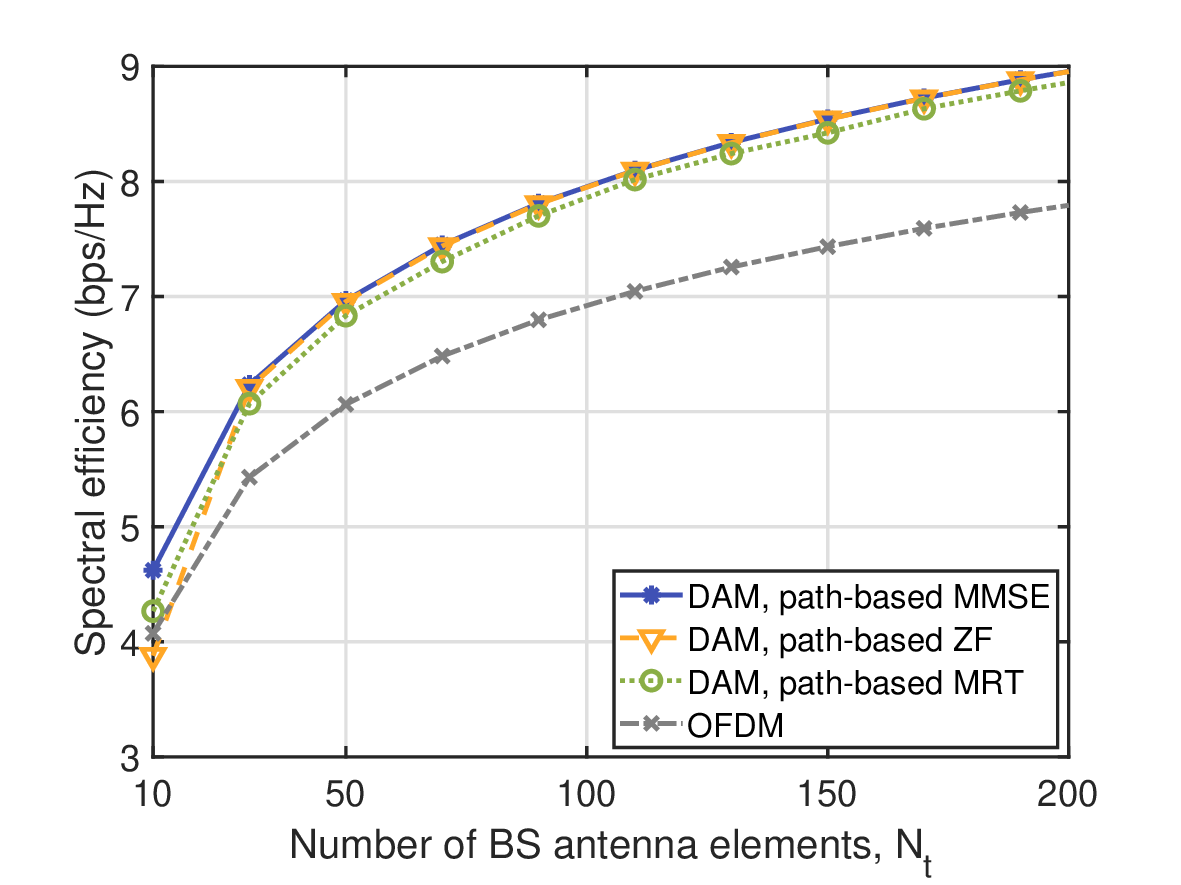}}
 \caption{Spectral efficiency versus the number of BS antenna elements for OFDM and path-based MMSE, ZF, and MRT DAM beamforming schemes.}
 \label{spectralEfficiencyVersusAntennaNumber}
 \end{figure}
 Fig.~\ref{spectralEfficiencyVersusAntennaNumber} shows the spectral efficiency versus antenna number $N_t$ for the proposed path-based MMSE, ZF, and MRT DAM beamforming schemes, respectively, together with the benchmarking OFDM scheme. The number of reflecting elements for each IRS is $M = {M_h}{M_v} = 256$, with ${M_h} = {M_v} = 16$. The transmit power is $P =40$ dBm. It is observed that for relatively small antenna number, e.g., $N_t = 10$, path-based MRT beamforming slightly outperforms path-based ZF beamforming. This is because path-based ZF beamforming sacrifices part of spatial dimensions for eliminating the ISI completely. As $N_t$ increases, the three DAM beamforming schemes give quite similar performance, especially when $N_t$ is large. This is expected since when the number of BS antennas $N_t$ is much larger than the number of IRSs $L$, the three DAM beamforming schemes are capable of well separating all the multi-path signal components in space, and hence the ISI can be effectively mitigated. Such a result is quite attractive for large antenna array of BSs in 6G communications, for which the performance of the low-complexity MRT beamforming approaches to that of the optimal MMSE beamforming scheme. It is also observed that when $N_t$ is large, the DAM beamforming schemes significantly outperform the conventional OFDM scheme, which is mainly attributed to the saving of guard interval. Specifically, the overhead of OFDM is ${n_{{\rm{OFDM}}}}{{\tilde n}_{\max }}/{n_c} = \frac{{217 \times 77}}{{1.28 \times {{10}^5}}} = 13.05\% $, whereas that of DAM is only $2{{\tilde n}_{\max }}/{n_c} = \frac{{154}}{{1.28 \times {{10}^5}}} = 0.12\% $.

 \begin{figure}[!t]
 \centering
 \centerline{\includegraphics[width=3.5in,height=2.625in]{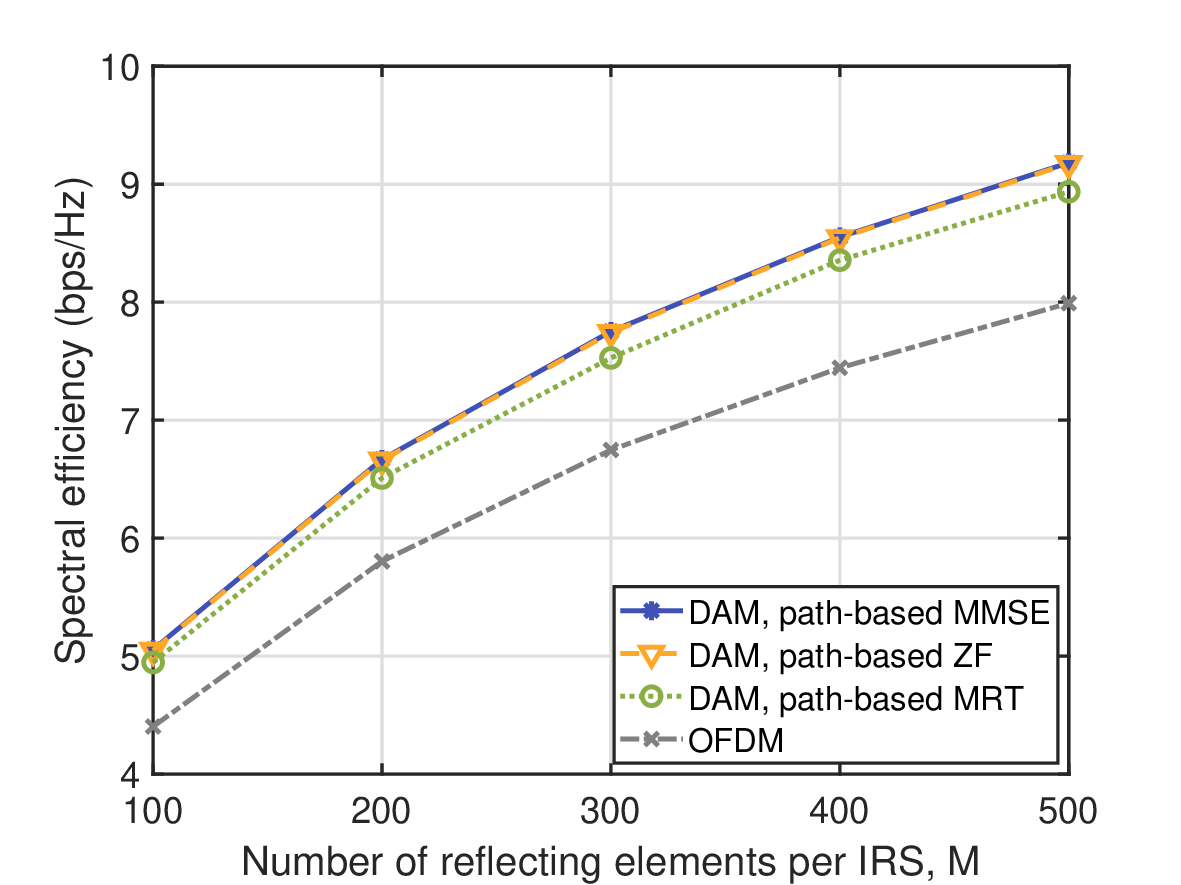}}
 \caption{Spectral efficiency versus the number of reflecting elements per IRS for OFDM and path-based MMSE, ZF, and MRT DAM beamforming schemes.}
 \label{spectralEfficiencyVersusReflectingNumber}
 \end{figure}
 Fig.~\ref{spectralEfficiencyVersusReflectingNumber} shows the spectral efficiency versus the number of reflecting elements $M$ for the proposed path-based MMSE, ZF, and MRT DAM beamforming schemes and the benchmarking OFDM scheme. The number of antennas is $N_t =64$, and the transmit power is $P=40$ dBm. The number of reflecting elements along the horizontal direction is fixed as $M_h = 10$, and $M$ is varied by varying the number of elements along the vertical direction $M_v$. It is observed that the spectral efficiency of all schemes increases with the number of reflecting elements, as expected. Furthermore, the proposed path-based MRT, MMSE and ZF DAM beamforming schemes all give comparable performance, which is expected due to the super spatial resolutions by the BS and IRSs' joint beamforming design. It is also observed that the three proposed DAM beamforming schemes significantly outperform the benchmarking OFDM scheme, thanks to its saving of the guard interval overhead.

 \begin{figure}[!t]
 \centering
 \centerline{\includegraphics[width=3.5in,height=2.625in]{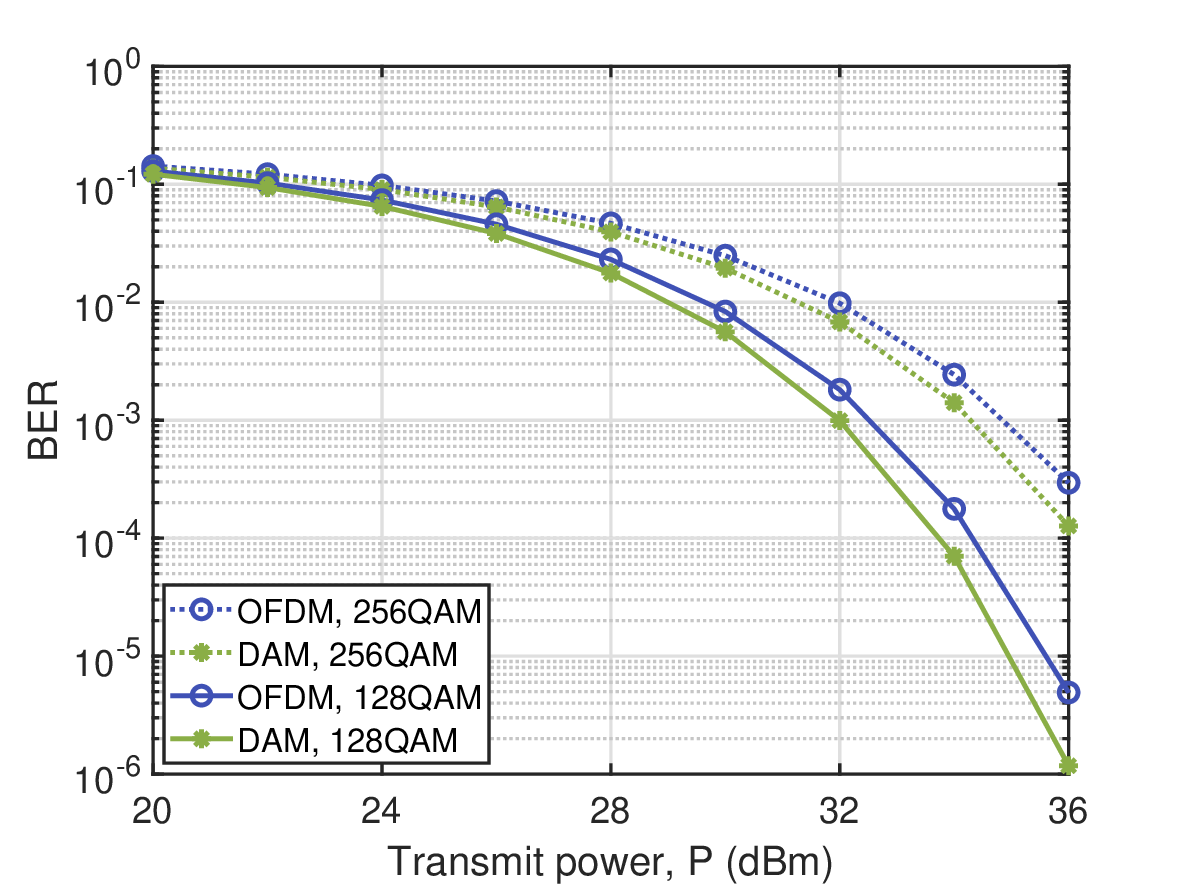}}
 \caption{BER comparison for the proposed path-based ZF DAM beamforming and the conventional OFDM.}
 \label{BERComparisionDAMVersusOFDM}
 \end{figure}
 Fig.~\ref{BERComparisionDAMVersusOFDM} shows the BER performance versus the transmit power $P$ for the proposed path-based ZF DAM beamforming and the benchmarking scheme of OFDM. Both 128 and 256 quadrature amplitude modulation (QAM) are considered. The number of antennas is ${N_t} = 128$, and that of reflecting elements for each IRS is $M = {M_h}{M_v} = 256$, with ${M_h} = {M_v} = 16$. Specifically, the BER expression of the OFDM-based multi-IRS aided communication can be approximated by \cite{xia2001precoded}
 \begin{equation}\label{OFDMBER}
 {P_{e,{\rm{OFDM}}}} = \frac{1}{K}\sum\limits_{k = 1}^{K} {{P_e}\left( {\frac{{{K}{p_k^{\star}}{{\left\| {{\bf{h}}\left[ k \right]} \right\|}^2}}}{{\left( {{K} + {{\tilde n}_{\max}}} \right){\sigma ^2}/{K}}}} \right)},
 \end{equation}
 where ${p_k^ \star }$ is obtained by the WF power allocation in Section \ref{sectionBenchmarkOFDMScheme}, and ${P_e}\left( \gamma \right)$ denotes the BER versus $\gamma$ over the AWGN channel. On the other hand, thanks to perfect delay alignment achieved by path-based ZF beamforming, the BER expression of DAM transmission is ${P_{e,{\rm{DAM}}}} = {P_e}\left( {{\gamma _{{\rm{ZF}}}}} \right)$. It is observed that the proposed DAM technique gives better BER performance than the conventional OFDM. This is expected since OFDM requires more CP overhead, which is reflected by the term $K/\left( {K + {{\tilde n}_{\max }}} \right)$ in \eqref{OFDMBER}.

 \begin{figure}[!t]
 \centering
 \centerline{\includegraphics[width=3.5in,height=2.625in]{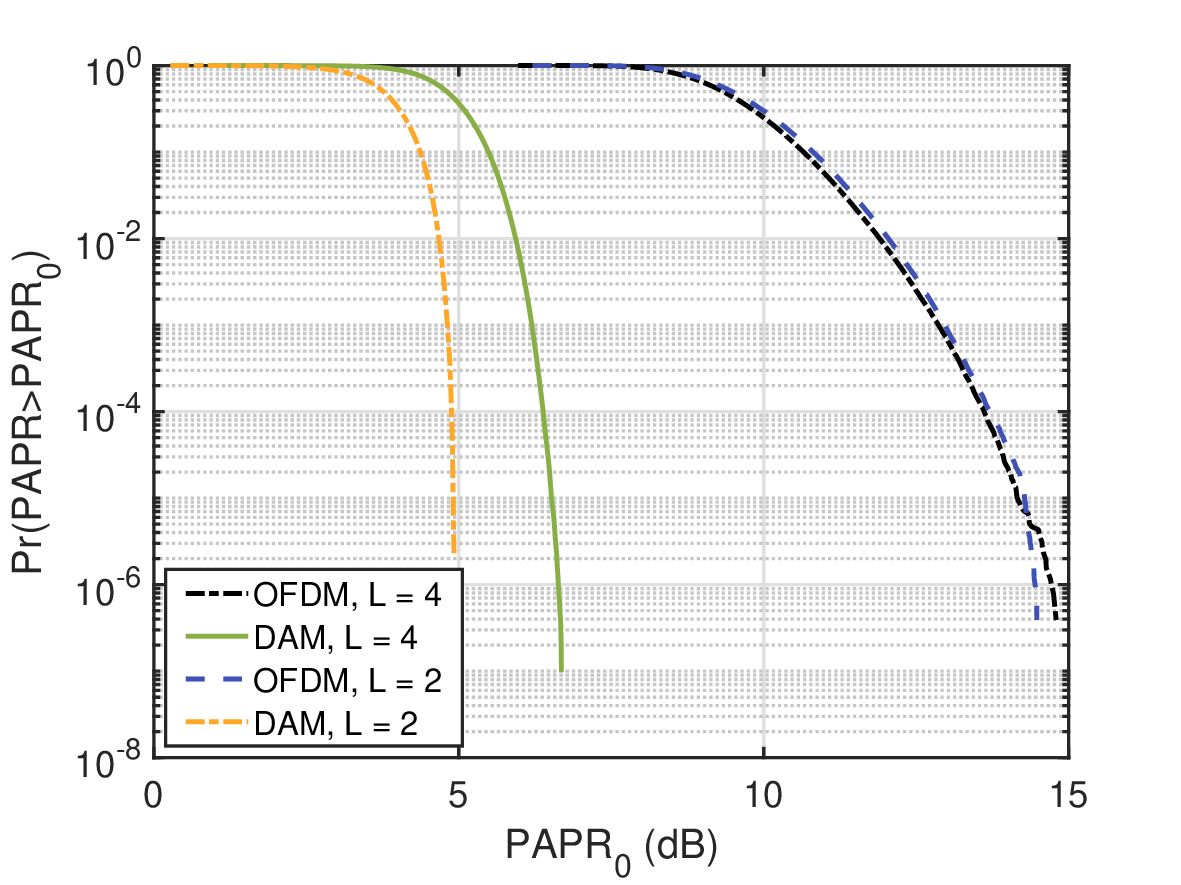}}
 \caption{PAPR comparison for the proposed path-based ZF DAM beamforming and the conventional OFDM.}
 \label{PAPRComparison}
 \end{figure}
 Last, Fig.~\ref{PAPRComparison} shows the PAPR comparison for the proposed path-based ZF DAM beamforming and OFDM, by using the 128QAM modulation. Specifically, the metric of complementary cumulative distribution function (CCDF) is used to evaluate the PAPR performance \cite{han2005overview}. The transmit power is $P=40$ dBm, and the number of antennas and reflecting elements for each IRS are the same as those of Fig.~\ref{BERComparisionDAMVersusOFDM}. For the case of $L=2$ IRSs, the first two IRS locations are used. It is observed that the DAM technique achieves significantly lower PAPR than the conventional OFDM, and the PAPR reduction is more pronounced as the number of IRSs $L$ decreases. This is expected since for OFDM, the signals from $K$ sub-carriers are superimposed on each antenna, whereas for DAM, only $L+1 \ll K$ multi-path signals are mixed on each antenna, as can be inferred from \eqref{transmitSignalDAM}.

\section{Conclusion}\label{sectionConclusion}
 This paper studied the multi-IRS aided communication with the single-carrier DAM transmission. It was shown that for the ideal case of asymptotically orthogonal channels with distinct delays, an ISI-free AWGN channel can be obtained with the simple path-based MRT beamforming at the BS. Besides, as long as the number of BS antennas is no smaller than the number of IRSs with significant multi-paths, ISI-free AWGN channel can still be obtained by applying path-based ZF beamforming tailored for DAM, without the need of sophisticated channel equalization or multi-carrier transmission. For the more general DAM transmission, the optimal path-based MMSE beamforming and the low-complexity path-based MRT beamforming were respectively studied, and it was shown that their performance is similar to that of the path-based ZF beamforming. Simulation results demonstrated that DAM outperforms OFDM in terms of spectral efficiency, BER, and PAPR.

 It is worth mentioning that despite its promising advantages, DAM also faces practical challenges. For example, DAM transmission requires CSI at the BS to implement delay pre-compensation and path-based beamforming. Thus, the efficient channel estimation method tailored for DAM and the performance characterization under CSI errors are worthy of further investigation. For ease of practical implementation, the quasi-static phase shift design based on statistical CSI \cite{han2019large,jia2022robust}, and low-complexity path-based beamforming design deserve further studies. Besides, the extensions of the results in this paper to more general channel models and multi-user systems are important to investigate in the future work.

\begin{appendices}
\section{Proof of Theorem \ref{PathBasedZFBeamformingTheorem}}\label{proofOfPathBasedZFBeamformingTheorem}
 By defining ${{\bar \mu }_{l'}} = {\mu _{l'}}{\left\| {{{\bf{w}}_{l'}}} \right\|^2}$, problem \eqref{subProblem2EquivalentProblem} can be equivalently written as
 \begin{equation}\label{subProblem3EquivalentProblem}
 \begin{aligned}
 \mathop {\max }\limits_{\left\{ {{\bar \mu} _{l'}} \right\}_{l' = 0}^L} &\ \ \frac{1}{{{\sigma ^2}}}{\left| {\sum\limits_{l' = 0}^L {\frac{{\sqrt {{{\bar \mu }_{l'}}} }}{{\left\| {{{\bf{w}}_{l'}}} \right\|}}} } \right|^2}\\
 {\rm{s.t.}} &\ \ \sum\limits_{l' = 0}^L {{{\bar \mu }_{l'}}}  \le P.
 \end{aligned}
 \end{equation}
 According to the Cauchy-Schwarz inequality, the objective function in \eqref{subProblem3EquivalentProblem} satisfies
 \begin{equation}\label{DAMReceivedSNRCauchy-Schwarzinequality}
 \begin{aligned}
 \frac{1}{\sigma ^2}{\left| {\sum\limits_{l' = 0}^L {\frac{{\sqrt {{{\bar \mu} _{l'}}} }}{{\left\| {{{\bf{w}}_{l' }}} \right\|}}} } \right|^2} \le \frac{1}{\sigma ^2}\left( {\sum\limits_{l' = 0}^L {{{\bar \mu} _{l'}}} } \right)\left( {\sum\limits_{l' = 0}^L {\frac{1}{{{{\left\| {{{\bf{w}}_{l' }}} \right\|}^2}}}} } \right),
 \end{aligned}
 \end{equation}
 where the equality holds if and only if $\sqrt {{{\bar \mu} _{l'}}}  = \lambda /\left\| {{{\bf{w}}_{l'}}} \right\|$ for some $\lambda$, $\forall l'$. With the power constraint, we have
 \begin{equation}
 \sum\limits_{l' = 0}^L {{{\bar \mu }_{l'}}}  = {\lambda ^{\rm{2}}}\sum\limits_{l' = 0}^L {\frac{1}{{{{\left\| {{{\bf{w}}_{l'}}} \right\|}^2}}}}  \le P.
 \end{equation}
 Therefore, an optimal $\lambda$ to maximize the objective function is
 \begin{equation}\label{optimalConstantlambda}
 {\lambda ^ \star } = \sqrt {\frac{P}{{\sum\limits_{i = 0}^L {\frac{1}{{{{\left\| {{{\bf{w}}_{i }}} \right\|}^2}}}} }}}.
 \end{equation}
 Then the optimal power allocation coefficient to \eqref{subProblem2EquivalentProblem} is
 \begin{equation}\label{optimalPowerAllocationCoeffZFProof}
 \mu _{l'}^ \star  = \frac{{\bar \mu _{l'}^ \star }}{{{{\left\| {{{\bf{w}}_{l'}}} \right\|}^2}}} = \frac{{{{\left( {{\lambda ^ \star }} \right)}^2}}}{{{{\left\| {{{\bf{w}}_{l'}}} \right\|}^4}}} = \frac{P}{{\sum\limits_{i = 0}^L {\frac{1}{{{{\left\| {{{\bf{w}}_i}} \right\|}^2}}}} }}\frac{1}{{{{\left\| {{{\bf{w}}_{l'}}} \right\|}^4}}}, \ \forall l',
 \end{equation}
 and the optimal beamforming vectors for (P-ZF1) are
 \begin{equation}\label{optimalTransmitBeamformingProof}
 {\bf{f}}_{l'}^ \star  = \sqrt {\mu _{l'}^ \star } {{\bf{w}}_{l'}}
 = \frac{{\sqrt P {{\bf{w}}_{l'}}}}{{\sqrt {\sum\limits_{i = 0}^L {\frac{1}{{{{\left\| {{{\bf{w}}_{i}}} \right\|}^2}}}} } {{\left\| {{{\bf{w}}_{l'}}} \right\|}^2}}}, \ \forall l'.
\end{equation}
 By substituting \eqref{optimalTransmitBeamformingProof} into the objective function of (P-ZF1), the resulting SNR in \eqref{optimalTransmitBeamformingSNR} can be obtained. The proof of Theorem \ref{PathBasedZFBeamformingTheorem} is thus completed.

\section{Proof of Proposition \ref{PathBasedZFBeamformingConvergence}}\label{proofOfPathBasedZFBeamformingConvergence}
 Let $\{ {\bf{f}}_{l'}^{k}\} _{l' = 0}^{L}$ and $\{ {{\bf{v}}_{l}^k} \}_{l = 1}^{L}$ be the resulting optimization variables after the $k$th iteration. Furthermore, let $\gamma \left( {\{ {\bf{f}}_{l'}^k\} _{l' = 0}^L,\{ {\bf{v}}_l^k\} _{l = 1}^L} \right)$ and $\gamma _{\rm{lb}}^{\rm{phase}}\left(\{ {\bf{f}}_{l'}^k\} _{l' = 0}^L,\{ {\bf{v}}_l^k\} _{l = 1}^L\right)$ denote the objective value of (P-ZF) and problem \eqref{subProblem1EquivalentProblem3}, respectively. At the $k$th iteration, for step 3 in Algorithm~\ref{alg1}, since the optimal solution to (P-ZF1) is obtained for given $\{ {{\bf{v}}_{l}^k} \}_{l = 1}^{L}$, we have
 \begin{equation}\label{ZFConvergencePathBeamforming}
 \gamma \left( {\left\{ {{\bf{f}}_{l'}^k} \right\}_{l' = 0}^L,\left\{ {{\bf{v}}_l^k} \right\}_{l = 1}^L} \right) \le \gamma \left( {\left\{ {{\bf{f}}_{l'}^{k + 1}} \right\}_{l' = 0}^L,\left\{ {{\bf{v}}_l^k} \right\}_{l = 1}^L} \right).
 \end{equation}
 For given $\{ {\bf{f}}_{l'}^{k+1}\} _{l' = 0}^{L}$ and $\{ {{\bf{v}}_{l}^k} \}_{l = 1}^{L}$ in step 4, we have
 \begin{equation}\label{ZFConvergencePhaseShift}
 \begin{aligned}
 & \gamma \left( {\left\{ {{\bf{f}}_{l'}^{k + 1}} \right\}_{l' = 0}^L,\left\{ {{\bf{v}}_l^k} \right\}_{l = 1}^L} \right) \mathop  = \limits^{(a)} \gamma _{{\rm{lb}}}^{{\rm{phase}}}\left( {\left\{ {{\bf{f}}_{l'}^{k + 1}} \right\}_{l' = 0}^L,\left\{ {{\bf{v}}_l^k} \right\}_{l = 1}^L} \right)\\
 &\ \ \ \ \ \ \ \ \ \ \ \ \ \ \ \ \ \ \ \ \ \ \ \ \ \ \ \ \mathop  \le \limits^{\left( b \right)} \gamma _{{\rm{lb}}}^{{\rm{phase}}}\left( {\left\{ {{\bf{f}}_{l'}^{k + 1}} \right\}_{l' = 0}^L,\left\{ {{\bf{v}}_l^{k + 1}} \right\}_{l = 1}^L} \right)\\
 &\ \ \ \ \ \ \ \ \ \ \ \ \ \ \ \ \ \ \ \ \ \ \ \ \ \ \ \ \mathop  \le \limits^{\left( c \right)} \gamma \left( {\left\{ {{\bf{f}}_{l'}^{k + 1}} \right\}_{l' = 0}^L,\left\{ {{\bf{v}}_l^{k + 1}} \right\}_{l = 1}^L} \right),
 \end{aligned}
 \end{equation}
 where the equality $\left(a\right)$ holds since the first-order Taylor approximation in \eqref{taylorApproximationPhaseShift} is tight at the given local point $\{ {{\bf{v}}_{l}^k} \}_{l = 1}^{L}$, which means that problem \eqref{subProblem1EquivalentProblem2} at $\{ {{\bf{v}}_{l}^k} \}_{l = 1}^{L}$ has the same objective value as that of problem \eqref{subProblem1EquivalentProblem3}; $\left( b \right)$ holds since in step 4, problem \eqref{subProblem1EquivalentProblem3} is solved via SCA that yields a non-decreasing objective value; $\left( c \right)$ holds since the objective value of \eqref{subProblem1EquivalentProblem3} is the lower bound of that of its original problem \eqref{subProblem1EquivalentProblem2} at $\{ {{\bf{v}}_{l}^{k+1}} \}_{l = 1}^{L}$. It is observed from \eqref{ZFConvergencePhaseShift} that although only an approximate optimization problem \eqref{subProblem1EquivalentProblem3} is solved for obtaining the phase shifts, the objective value of (P-ZF2) is still non-decreasing after each iteration. Moreover, with \eqref{ZFConvergencePathBeamforming} and \eqref{ZFConvergencePhaseShift}, we have
 \begin{equation}\label{ZFConvergenceOverall}
 \gamma \left( {\left\{ {{\bf{f}}_{l'}^k} \right\}_{l' = 0}^L,\left\{ {{\bf{v}}_l^k} \right\}_{l = 1}^L} \right) \le \gamma \left( {\left\{ {{\bf{f}}_{l'}^{k + 1}} \right\}_{l' = 0}^L,\left\{ {{\bf{v}}_l^{k+1}} \right\}_{l = 1}^L} \right),
 \end{equation}
 which indicates that the objective value of (P-ZF) is non-decreasing over each iteration. Thus, Algorithm~\ref{alg1} is guaranteed to converge, and Proposition \ref{PathBasedZFBeamformingConvergence} is proved.
\end{appendices}


\bibliographystyle{IEEEtran}
\bibliography{refDAMIRS}

\end{document}